\documentclass[preprint,twocolumn,3p]{elsarticle}

\usepackage{lineno,hyperref}
\modulolinenumbers[5]

\usepackage[utf8]{inputenc}
\usepackage{amsmath}
\usepackage{natbib}
\usepackage{graphicx}
\usepackage{gensymb}
\usepackage{url}
\usepackage{mathtools}
\usepackage[varg]{txfonts}
\usepackage{textcomp}
\usepackage[super]{nth}

\DeclarePairedDelimiter{\ceil}{\lceil}{\rceil}

\renewcommand{\vec}[1]{\mathbf{#1}}

\biboptions{authoryear}

\begin{document}

\begin{frontmatter}

\title{Generating realistic synthetic meteoroid orbits}

\author[uwoes,uwopa]{Denis Vida}
\ead{dvida@uwo.ca}

\author[uwopa]{Peter G. Brown}
\author[uwopa]{Margaret Campbell-Brown}

\address[uwoes]{Department of Earth Sciences, University of Western Ontario, London, Ontario, N6A 5B7, Canada}
\address[uwopa]{Department of Physics and Astronomy, University of Western Ontario, London, Ontario, N6A 3K7, Canada}

\begin{abstract}

\textit{Context.} Generating a synthetic dataset of meteoroid orbits is a crucial step in analysing the probabilities of random grouping of meteoroid orbits in automated meteor shower surveys. Recent works have shown the importance of choosing a low similarity threshold value of meteoroid orbits, some pointing out that the recent meteor shower surveys produced false positives due to similarity thresholds which were too high. On the other hand, the methods of synthetic meteoroid orbit generation introduce additional biases into the data, thus making the final decision on an appropriate threshold value uncertain.

\textit{Aims.} As a part of the ongoing effort to determine the nature of meteor showers and improve automated methods, it was decided to tackle the problem of synthetic meteoroid orbit generation, the main goal being to reproduce the underlying structure and the statistics of the observed data in the synthetic orbits.

\textit{Methods.} A new method of generating synthetic meteoroid orbits using the Kernel Density Estimation method is presented. Several types of approaches are recommended, depending on whether one strives to preserve the data structure, the data statistics or to have a compromise between the two.

\textit{Results.} The improvements over the existing methods of synthetic orbit generation are demonstrated. The comparison between the previous and newly developed methods are given, as well as the visualization tools one can use to estimate the influence of different input parameters on the final data.

\end{abstract}


\end{frontmatter}


\section{Introduction}
Identification of small solar system bodies having a common origin is crucial to understanding the broader evolution of the solar system. Examples of such populations include Kuiper-belt objects, main-belt asteroid families, sungrazing comets and meteoroid streams. In the latter example, a major roadblock to a common association is the problem of false positives, i.e. discriminating meteoroids having similar orbits due to common parentage from simple contamination. Recently, the problem of confusion due to false-positive orbital associations has been examined in the closely related topic of near--Earth asteroid (NEA) families by \cite{Schunova2012}. Using a model orbital distribution of the expected NEA population, they showed conclusively that no significant NEA ``streams" exist. A similar model-based approach to the statistical significance of meteoroid stream orbit associations has been harder to implement, due to the lack of a similarly detailed model of the unbiased ``true" meteoroid orbit distribution.

In recent years there has been a significant increase in the number of measured video meteoroid orbits. Among these are the Japanese SonotaCo Network\footnote{SonotaCo meteor orbit databases, \url{http://sonotaco.jp/doc/SNM/}} alone contributed 227,579 orbits, collected in the 2007 - 2015 period. The Cameras for Allsky Meteor Surveillance (CAMS) project has released more than 100,000 measured orbits collected to the end of 2013 \citep{jenniskens2016established}. The Croatian Meteor Network has published 39,891 orbits collected in the 2007 - 2013 period \citep{korlevic2013croatian}. Finally, EDMOND 5.0, a joint meteor orbit database of several European networks, contains 145,830 orbits \citep{rudawska2015independent}, which are also publicly available. With large number statistics for good quality meteoroid orbits now available, a natural question is how best to distinguish true meteor showers from statistical fluctuations within each dataset.

Searches for meteor showers among these datasets have been attempted \citep{rudawska2014new, jenniskens2016newly, andreic2014results, kornovs2014confirmation}. All such analyses have used one or more orbit dissimilarity criteria to gauge the similarity between individual orbits and thus establish groupings. Most of these orbit-based criteria are some variant of the original D-criterion developed by \cite{southworth1963statistics}. The D-criterion weights differences in orbital elements to establish proximity in orbital phase space. Variants on the original D-criterion have been proposed by \cite{drummond1981test}; \cite{jopek1993remarks}; and \cite{valsecchi1999meteoroid} among others. The challenge in using any D-criterion approach is choosing a threshold value below which two orbits are declared ``similar". In \cite{jenniskens2016cams} the authors point out that the results of automated meteor shower searches with fixed D-criteria thresholds are often unreliable, producing false groupings in orbits. In particular, because the sporadic environment is non-uniform, the probability of spurious groupings varies in different parts of the orbital phase space and among different surveys having different biases. Identification of weaker showers, in particular, fundamentally requires adoption of a background model to establish the false-positive rate to provide an objective metric for choice of threshold, independent of the clustering metric employed. 

Another approach for determining D-criteria threshold values was proposed by \citet{moorhead2016performance}, where synthetic ``showers" were generated instead of the synthetic sporadic background. A shower analogue for each of 30 analyzed showers was created: each analogue was similarly positioned in the Sun-centered ecliptic coordinate as the analyzed shower, but was offset in solar longitude from 60\degree to 300\degree, in increments of 10\degree. All orbital elements were computed for each analogue, and were separately inserted into the observed data. As it was known which orbit was inserted, and which was not, upon performing shower extraction with different threshold values of D-criteria it was possible to determine the false positive intrusion for each. Nevertheless, seasonal variations in the sporadic background and contributions from nearby showers were not accounted for, but an attempt was made to counter the influence of strong showers by iteratively removing showers from the dataset throughout the analysis, from strongest to weakest.

\cite{jopek2016probability} have recently examined the question of false-positives in the pairing of two meteoroid orbits. Their results showed that, for the datasets used to have at least $3\sigma$ confidence that the pairing was not random, one needs to use a smaller threshold than those usually assumed in automated searches. One step in their analysis was generating a synthetic meteor sporadic background from a known dataset, which was the basis for the later part of the analysis. Of their examined approaches, the most successful method they termed ``method E". This consisted of drawing random samples from histograms of individual orbital parameters ($e$, $\omega$, $\Omega$ and $i$) using the cumulative probability distribution (CPD) inversion method. The method was designed to reproduce the following properties of the input data: 

\renewcommand{\theenumi}{\alph{enumi}}
\begin{enumerate}
	\item  The fraction of meteors in each Monte Carlo run with positive ecliptic latitudes ($\beta > 0$) was calculated as $\beta_{fr}$ and was set to be the same as the observed dataset.
	
	\item The orbits were calculated such that they satisfy the Earth crossing condition:
	\begin{equation}\label{eq:earth_crossing}
	r = \frac{q(1+e)}{1 \pm e \cos{\omega}}
	\end{equation}
	where $r$ is the heliocentric distance of the meteoroid at the moment it collides with Earth (assumed to be fixed at $r = 1 [AU]$); $q$ is the perihelion distance; $e$ is the eccentricity and $\omega$ is the argument of perihelion. Note that the sign of $e \cos{\omega}$ is positive if the geocentric ecliptic latitude of the meteor radiant is negative. The sign was determined by generating a number $b$ from a uniform distribution $U(0,1)$, and if $b > \beta_{fr}$, the sign was chosen to be negative.
	
	\item The semi-major axis was calculated from the known relation:
	
	\begin{equation} \label{eq:a_calc}
	\frac{1}{a} = \frac{1-e}{q}
	\end{equation}
	
    Monte Carlo picks which had a larger $1/a$ than the maximum value in the observed dataset $1/a_{max}$ were rejected.

\end{enumerate}

\subsection{Test of existing model}
In developing a model to quantitatively test for false-positive associations between orbits we begin by independently examining method E of \cite{jopek2016probability}. For our reference dataset defining the sporadic background we use the CAMS video meteor orbit database \citep{jenniskens2016established}. The database already has a meteor shower identified with each measured meteor (where appropriate). Here we use only orbits which were not linked to any showers by \citet{jenniskens2016established}. In that work, showers were identified manually by looking at Sun-centered ecliptic and longitude of perihelion versus inclination plots. When a shower was identified, its orbits were extracted using the \citet{jopek1993remarks} $D_h$ criterion. The $D_h$ threshold value for shower association was manually adjusted until it was subjectively determined that the shower is well separated from the sporadic background. In \citet{jenniskens2016verification}, this method failed to identify a number of minor shower detected in a previous automated search by \citet{rudawska2014new} on the same dataset.

By way of quality control we rejected meteor trajectories with a convergence angle (camera-meteor-camera angle) of less then 15$\degree$ or an error in the geocentric velocity of more than 10\%, and orbits with $e \ge 1$. Applying these quality filters left a total of $n = 58,090$ orbits from the original dataset of 110,257 orbits. For this observed dataset of $n$ events, histograms were then generated for all five orbital parameters, each with $k = 80$ bins, calculated according to the Rice Rule:

\begin{equation}
k = \ceil{2n^{1/3}}
\end{equation}
where $\ceil{x}$ is the ceiling function, i.e. a function which maps a real number $x$ to the least integer greater than or equal to $x$.

\begin{figure}
  \includegraphics[width=\linewidth]{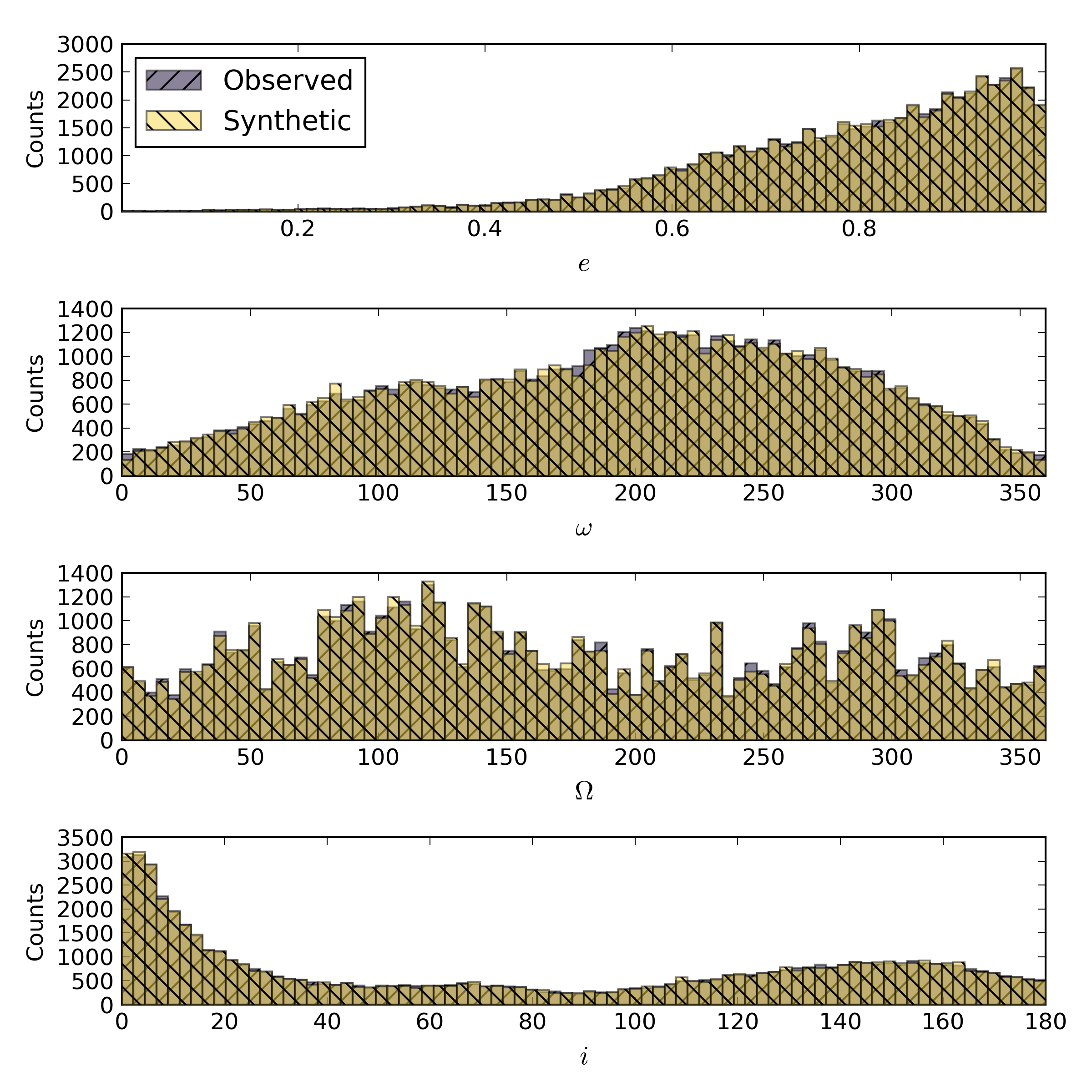}
  \caption{Histograms of all generated orbital parameters: $e$, $\omega$, $\Omega$ and $i$ by using the \cite{jopek2016probability} method E. Both the original (observed) and synthetic orbit histograms are shown.}
  \label{fig:all_elements_jopek}
\end{figure}

\begin{figure}
  \includegraphics[width=\linewidth]{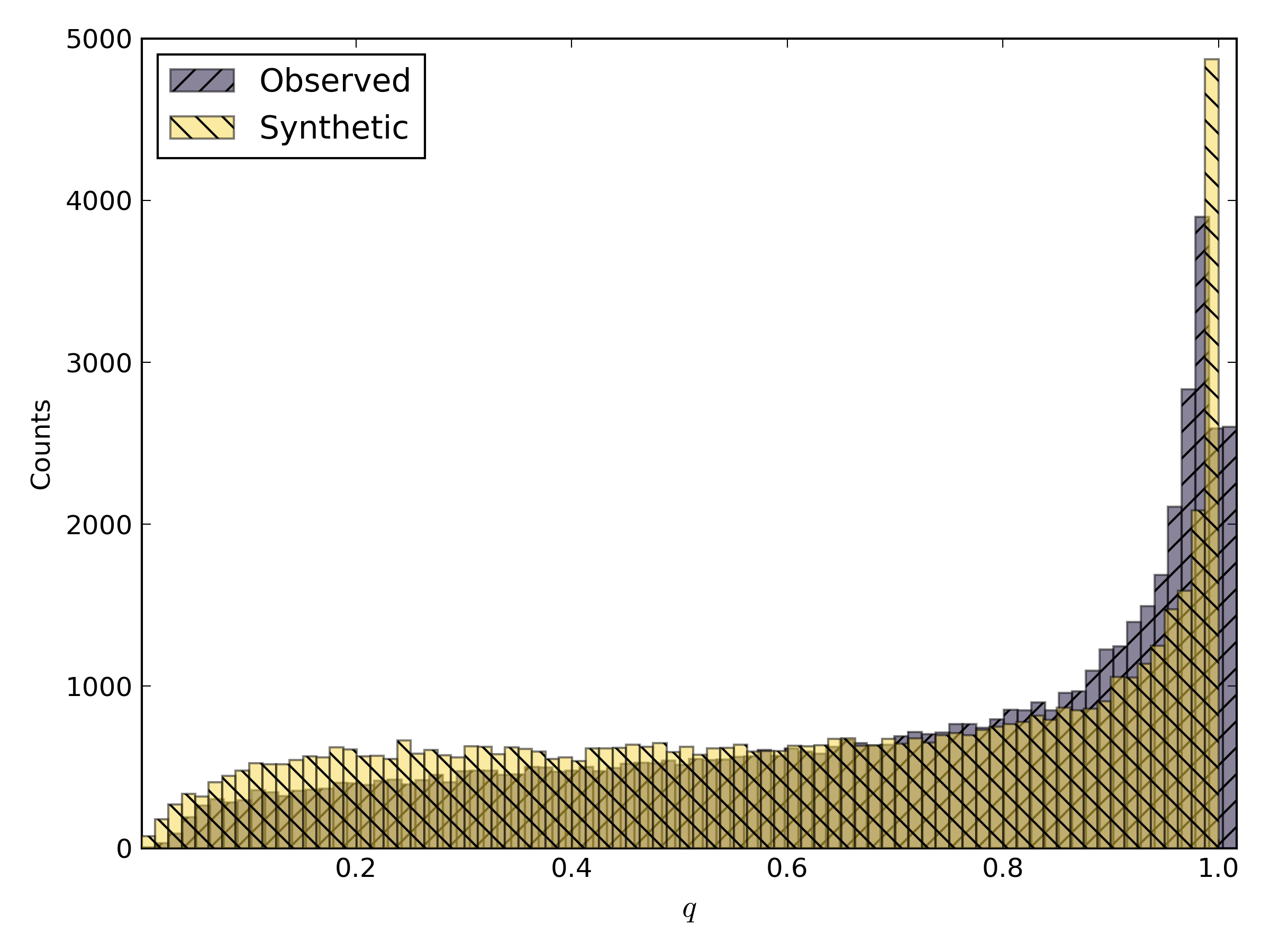}
  \caption{Comparison of the perihelion distance histogram of the synthetic orbits with the observed dataset. The synthetic orbits were generated by method E given in \cite{jopek2016probability}.}
  \label{fig:q_jopek}
\end{figure}

\begin{figure}
  \includegraphics[width=\linewidth]{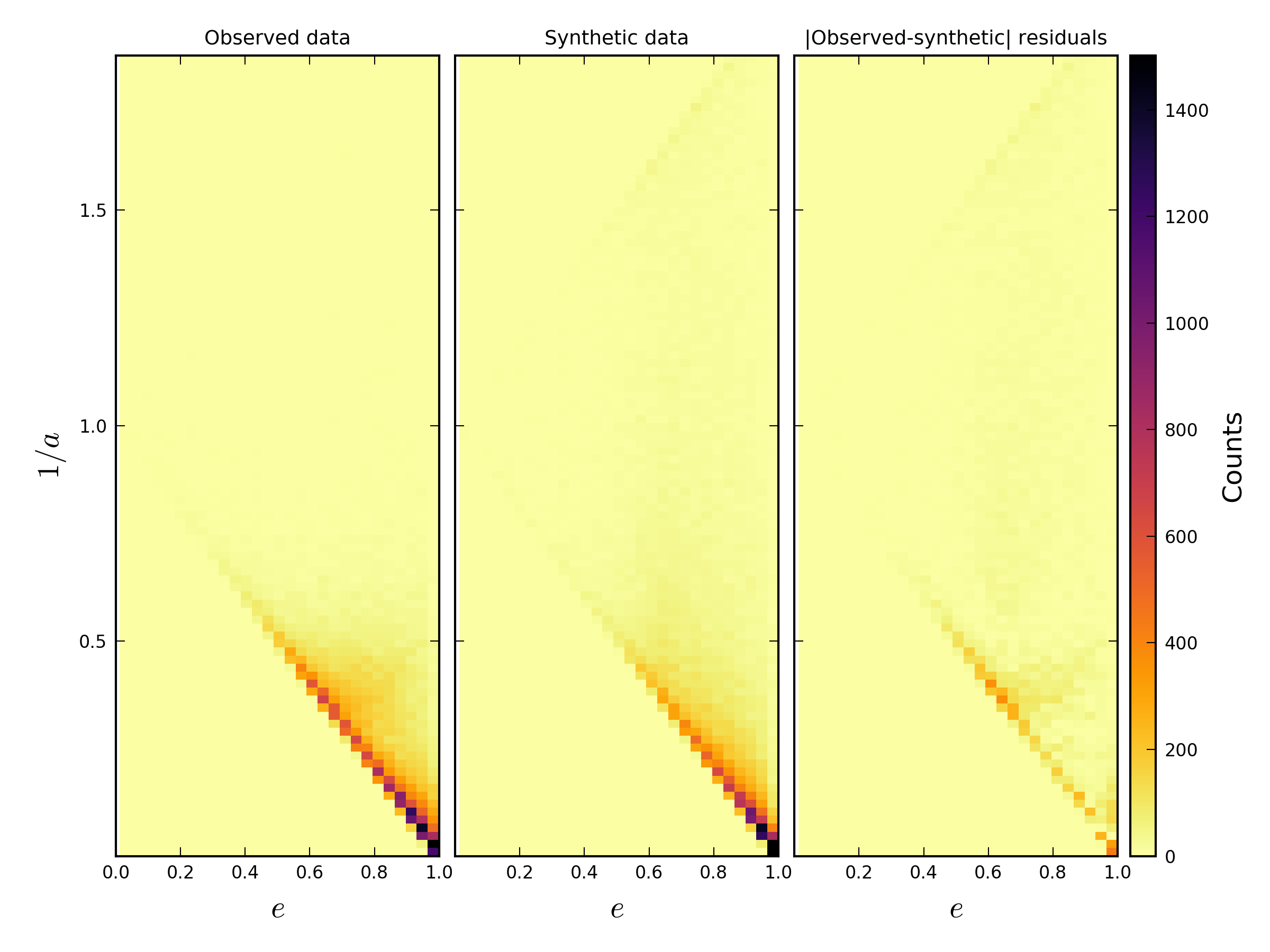}
  \caption{Inverse semi-major axis vs. eccentricity vs.  2D  histograms created by using the \cite{jopek2016probability} method E. The left inset shows a density plot of the observed data, the middle one shows the synthetic data, while the right inset shows absolute residuals after subtracting the synthetic data density plot from the observed data density plot.}
  \label{fig:e-1_a_jopek}
\end{figure}

\begin{figure}
  \includegraphics[width=\linewidth]{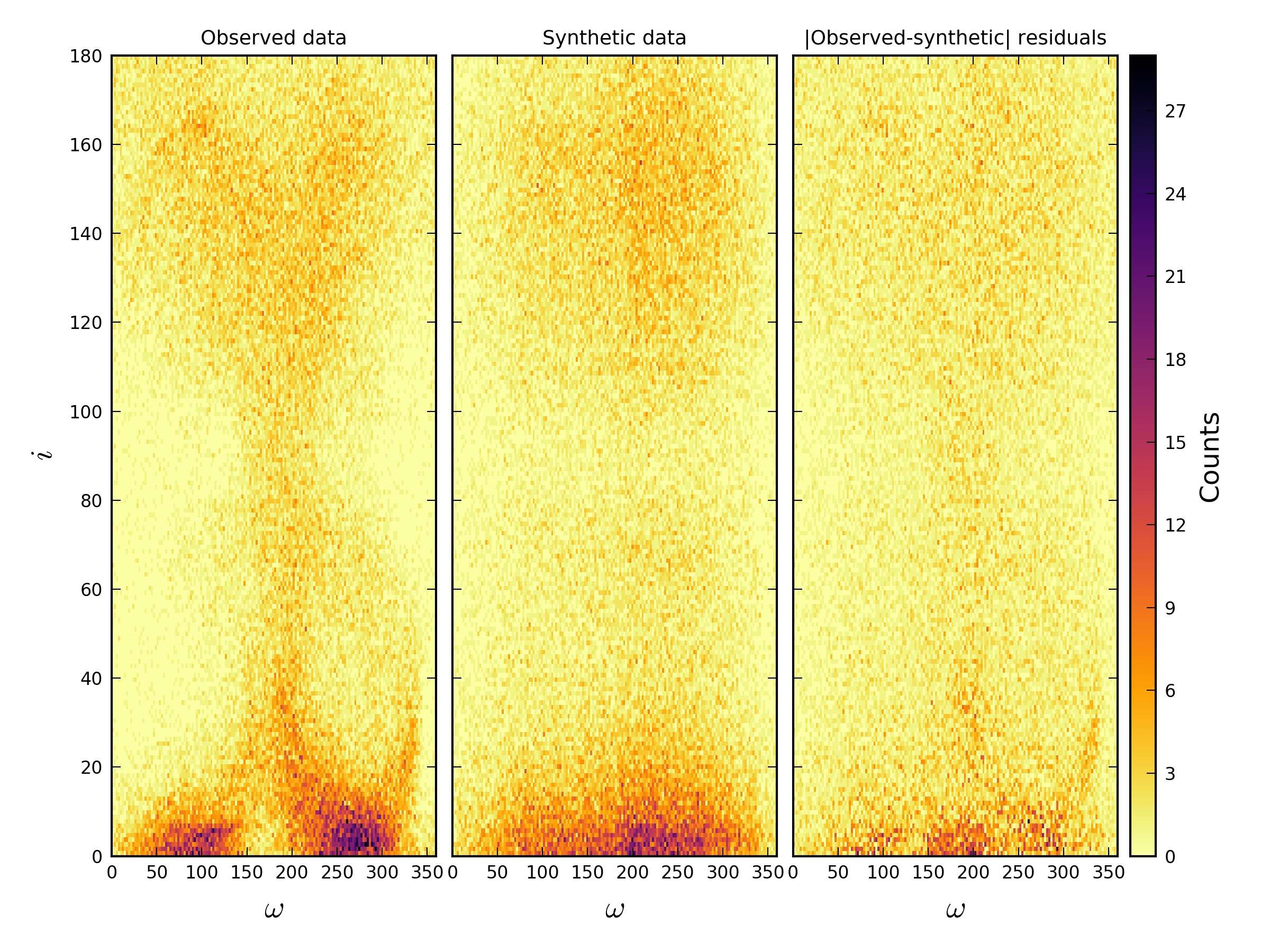}
  \caption{Argument of perihelion vs. inclination 2D histograms created by using the \cite{jopek2016probability} method E.}
  \label{fig:incl-peri_jopek}
\end{figure}

\begin{figure}
  \includegraphics[width=\linewidth]{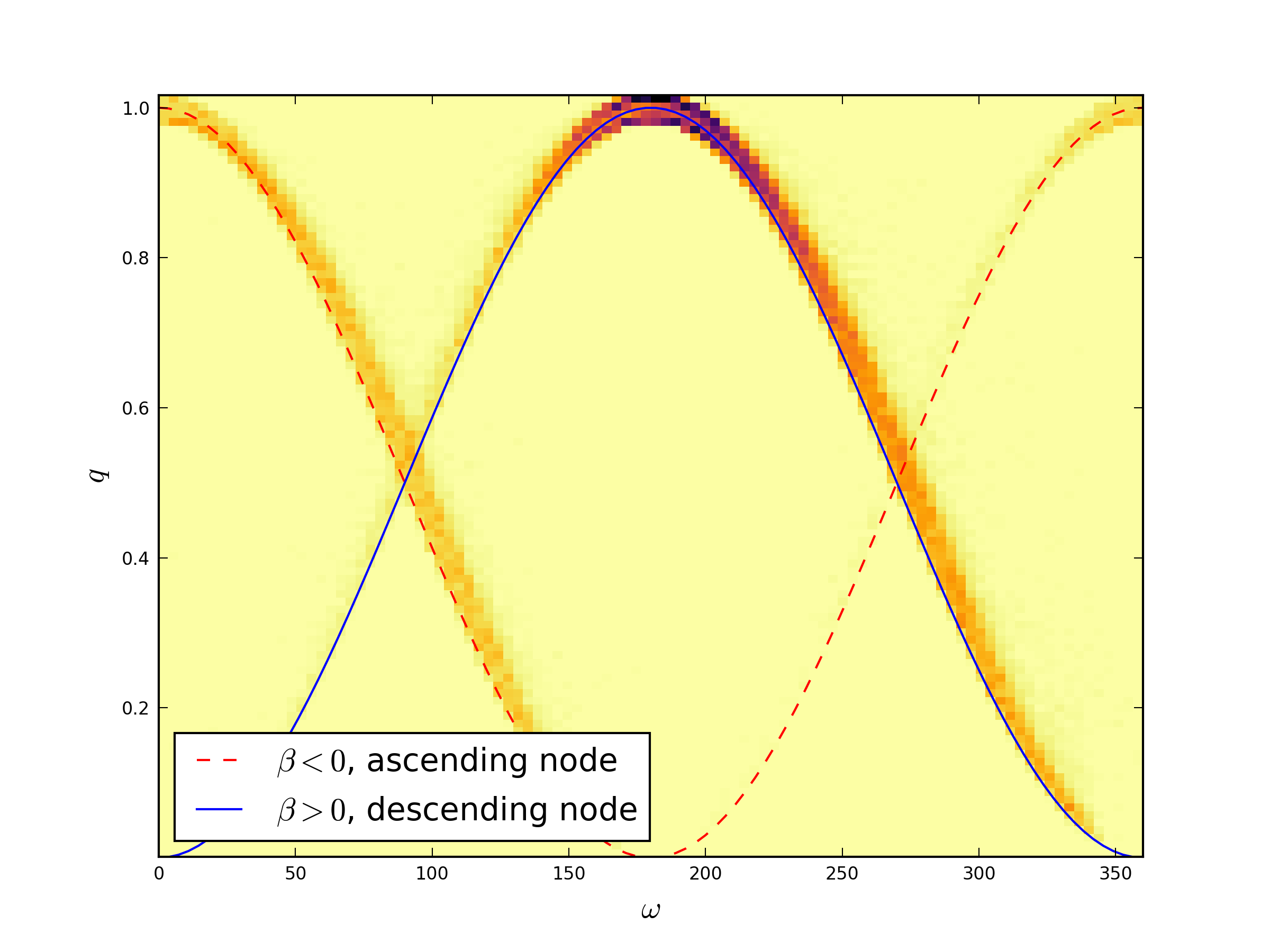}
  \caption{Argument of perihelion vs. perihelion distance 2D histogram for the observed data.}
  \label{fig:peri-q_explanation}
\end{figure}

\begin{figure}
  \includegraphics[width=\linewidth]{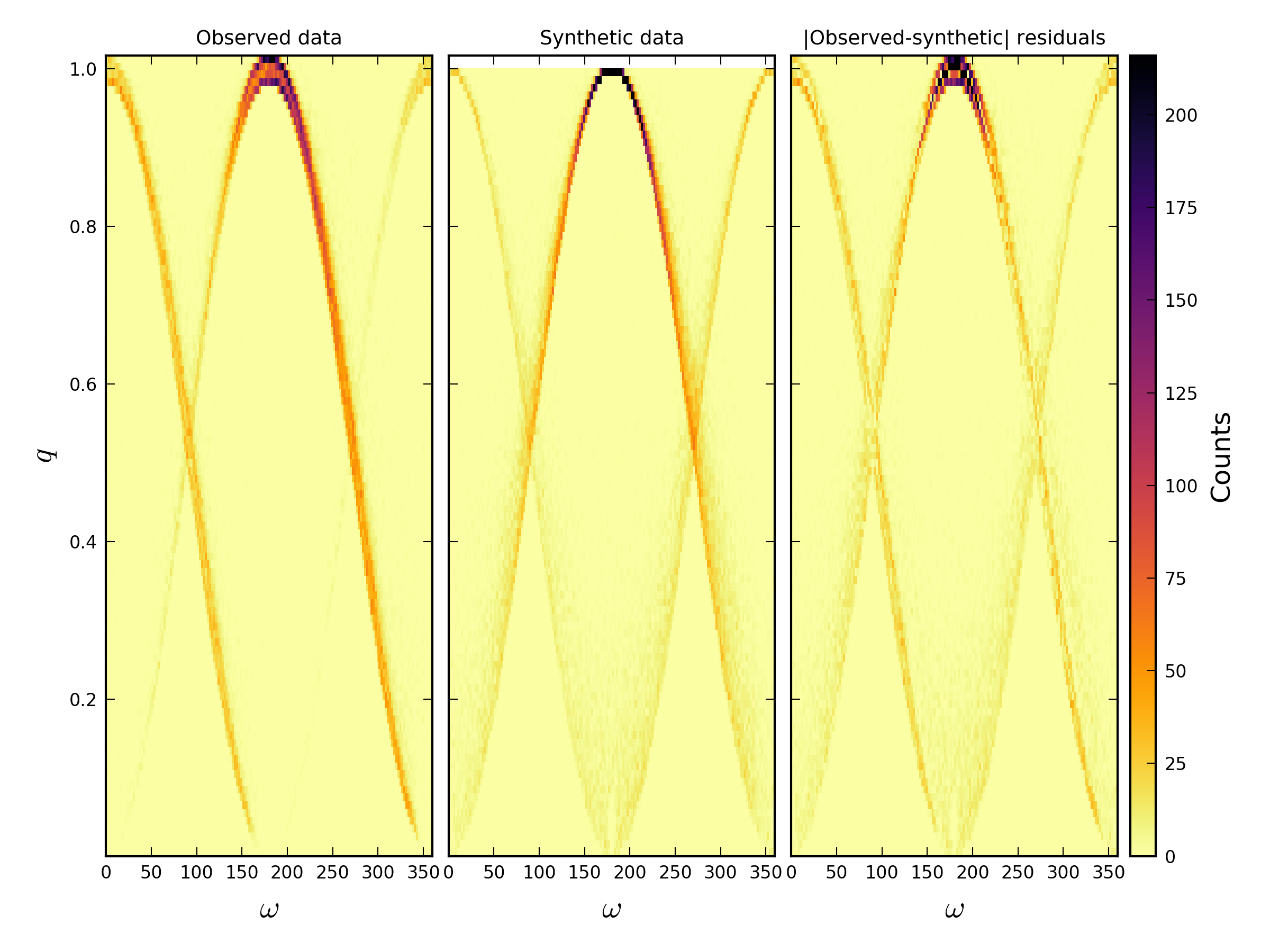}
  \caption{Argument of perihelion vs. perihelion distance 2D histograms created by using the \cite{jopek2016probability} method E.}
  \label{fig:peri-q_jopek}
\end{figure}

The samples were drawn following method E of \cite{jopek2016probability}. The results are shown in Figures \ref{fig:all_elements_jopek} to \ref{fig:peri-q_jopek}. While figures in this section show only one Monte Carlo run, multiple runs show the same general behaviour for all figures.

Figure \ref{fig:all_elements_jopek} shows histograms of both the observed and the generated synthetic orbits. The method leads to synthetic datasets which reproduce the character of the original (observed) histograms of orbital elements, without any major discrepancies between the two, the largest being approximately 100 counts in $\omega$.

Once these orbital parameters were synthetically generated, equation \ref{eq:earth_crossing} was used to calculate $q$. The results are shown in Figure \ref{fig:q_jopek}. Here a discrepancy is obvious: the observed data goes just beyond 1 AU, while the maximum perihelion distance of the synthetic orbits is at 1 AU. This caused the final synthetic bin to have a larger count then the observed bin, overestimating the number of orbits with $q \approx 1$. Furthermore, the synthetic data as produced by Method E suffer a general bias towards lower perihelion distances, their counts being systematically larger then those of observed data. A possible improvement might be realized by generating a different heliocentric distance of Earth, $r$, for each orbit, according to Earth's position at the time of encounter. We explored this approach but in the end chose not to use it as it does not fix other discrepancies found between the synthetic and observed orbital elements described below. 

As shown in the density plots in figures \ref{fig:e-1_a_jopek} to \ref{fig:peri-q_jopek}, additional discrepancies exist between the observed and synthetic data. Figure \ref{fig:e-1_a_jopek} shows a density plot of eccentricity and the inverse of the semi-major axis. The general correlations are preserved in the synthetic data, but the synthetic data are skewed towards higher values of eccentricity and semi-major axis compared to the observed dataset. Additionally, there are fewer synthetic orbits at $e \approx 0.6$, compared to the observed dataset.

Figure \ref{fig:incl-peri_jopek} shows a comparison of the correlation between the argument of perihelion and inclination for the observed and synthetic datasets.  The synthetic data plot is much more uniform than the observed dataset; it is not a good reproduction of the actual structure in the observed dataset as it fails to reproduce seasonal variations in sporadic sources, as well as producing a high amount of low-inclination orbits at $\omega \approx 180\degree$ which are not present in the observed data. Thus, the plot of absolute residuals between the two datasets shows numerous bins with high counts.

We have also found it important to take the $\omega$ vs. $q$ 2D histogram into consideration: figure \ref{fig:peri-q_explanation} shows the aforementioned plot for the observed data. The data exhibit a distinct structure with two main branches, of which the central one that peaks at $\omega \approx 180\degree$ is further divided into a high and a low $q$ branch of high density. The two main branches are:
\begin{enumerate}
    \item Orbits with ecliptic latitude $\beta < 0$, which met the Earth at or close to their ascending node, and which roughly follow $\frac{1}{2} \left( 1 + \cos \omega  \right)$.
    \item Orbits with ecliptic latitude $\beta > 0$, which met the Earth at or close to their descending node, and which roughly follow $\frac{1}{2} \left( 1 + \cos (\omega + \pi)  \right)$.
\end{enumerate}
Most of the meteors were encountered at their descending node and at $\omega \approx 180 \degree$, meaning that their perihelion distance will correspond to the distance of the Earth to the Sun. Because this distance is not constant throughout the Earth's orbit, the $q$ of meteor orbits at $\omega \approx 180 \degree$ varies accordingly. The larger number of meteors observed at aphelion and perihelion may be explained by observational biases, such the effective nightly sky coverage in different parts of the year (Figure 2 in \citet{jenniskens2016established}). Furthermore, the reason for the lack of orbits at $\beta < 0$, $\omega > 180\degree$ and $\beta > 0$, $\omega < 180 \degree$ is that the dataset does not contain any daytime orbits.

Finally, Figure \ref{fig:peri-q_jopek} shows a density plot of the argument of perihelion versus the perihelion distance for both synthetic and real data. The synthetic data on the other hand fail to quantitatively reproduce the left branch concentrating most orbits into a central branch at high values of perihelion distance. Moreover, there is no separation in the central branch into two denser high and low $q$ branches. This causes a very high difference between the two data sets, as visible on the residuals plot. Another difference is the lack of any orbits with $q > 1$, the cause of which has been previously discussed.

These discrepancies between the synthetic and observed datasets mean that Method E of \cite{jopek2016probability} cannot be confidently used to construct a synthetic sporadic background for CAMS data. With such differences one can only determine lower limits to the probability of false-postives for random pairings;  the real values are likely much higher.

In light of this result, we decided to develop a new method for generating synthetic orbits of meteors which better mimicks the characteristic correlations within an observed dataset and hence allows us to create a sporadic background model that is closer to the observed data. Note that our procedure does not replicate the true sporadic background, but rather the background as observed by a particular instrument/system and hence has that instrument's biases built into the model.

\section{Kernel density estimation overview}
In statistical analysis, one is often presented with a sample of independent random variables $X_1, X_2, ..., X_n$ which are distributed in n-dimensional parameter space according to some density $f$. To estimate $f$ from the sample, two classes of estimators can be employed: parametric and non-parametric. In the parametric approach, one assumes a fixed functional structure, an underlying model, whose parameters are approximated from the measurements. On the other hand, in the cases when the underlying model is unknown or too complex to parametrically define, non-parametric estimators are applied as they depend only on the data to reach a final estimation. The only assumptions made are that the density function exists and that it is differentiable.

The most popular non-parametric density estimator is the histogram, which is also often used as a data presentation device. To use this method, one specifies the origin $t_0$ and the width of individual bins $h$ in which the data points are accumulated. The density of a random variable $X$ is then represented as the height of individual bins. Despite its popularity and ease of use, the major drawback is the uncertainty of choosing the bin width which will correctly preserve the density information. For example, Sturges' rule gives an estimate of the number of bins to be used, while one needs to decide on the limits of the histogram to calculate the bin width. The rule was created to optimally work with normally distributed data, while the general recommendation for skewed or leptokurtotic data is to add more bins \citep{scott2015multivariate}. While there are other approaches for bin width estimation available, even for non-normal data, often the final number of bins is left as a subjective choice. Furthermore, histogram distributions are inherently not smooth and assume that all values in the same bin have the same frequency, while the true underlying distribution may not behave in such a discrete way.

To overcome most of the issues of histograms, kernel density estimators (KDEs), non-parametric in nature, are used. A good overview of the topic can be found in \cite{hwang1994nonparametric}. KDEs assume that each data point has a certain probability of appearance, which is represented as a kernel centred at each data point. If the kernels overlap, the overlapping probability is summed. The shape of the kernel is parametrically defined and can be arbitrarily chosen. In other words, kernel estimators smooth out the contribution of each point over its local neighbourhood. The only choice one is left with, apart from choosing the parametric curve representing the kernel, is its width, formally called the \textit{bandwidth}. If the data being handled are the product of a measurement, a Gaussian kernel can be assumed, and the bandwidth is then closely related to the standard deviation of the measurement. While the standard deviation (or a covariance matrix) can give a proper shape for the kernel, choosing an appropriate bandwidth is more problematic. For a 1D case, the KDE is given as follows:\\

\begin{equation} \label{eq:1d_ke}
\hat{f}(x) = \frac{1}{nh} \sum\limits_{i=1}^{n} K \left( \frac{x-x_i}{h} \right)
\end{equation}

where $x$ is the independent random variable, $x_i$ individual data points, $n$ the total number of data points, $K$ the kernel and $h$ the bandwidth. The kernel must have the following properties:

\begin{eqnarray}
K(x) \geq 0 \\
\int_{-\infty}^{\infty}K(t)dt = 1 \\
K(x) = K(-x)
\end{eqnarray}

the last equation means that the kernel function is symmetric with respect to the origin. From equation \ref{eq:1d_ke} we can conclude that the bandwidth $h$ directly determines the scale of the net influence of individual points on the result; thus, choosing this parameter is the most important step in the analysis. A review of methods for bandwidth estimation is given in \cite{jones1996brief}.

\begin{figure}
  \includegraphics[width=\linewidth]{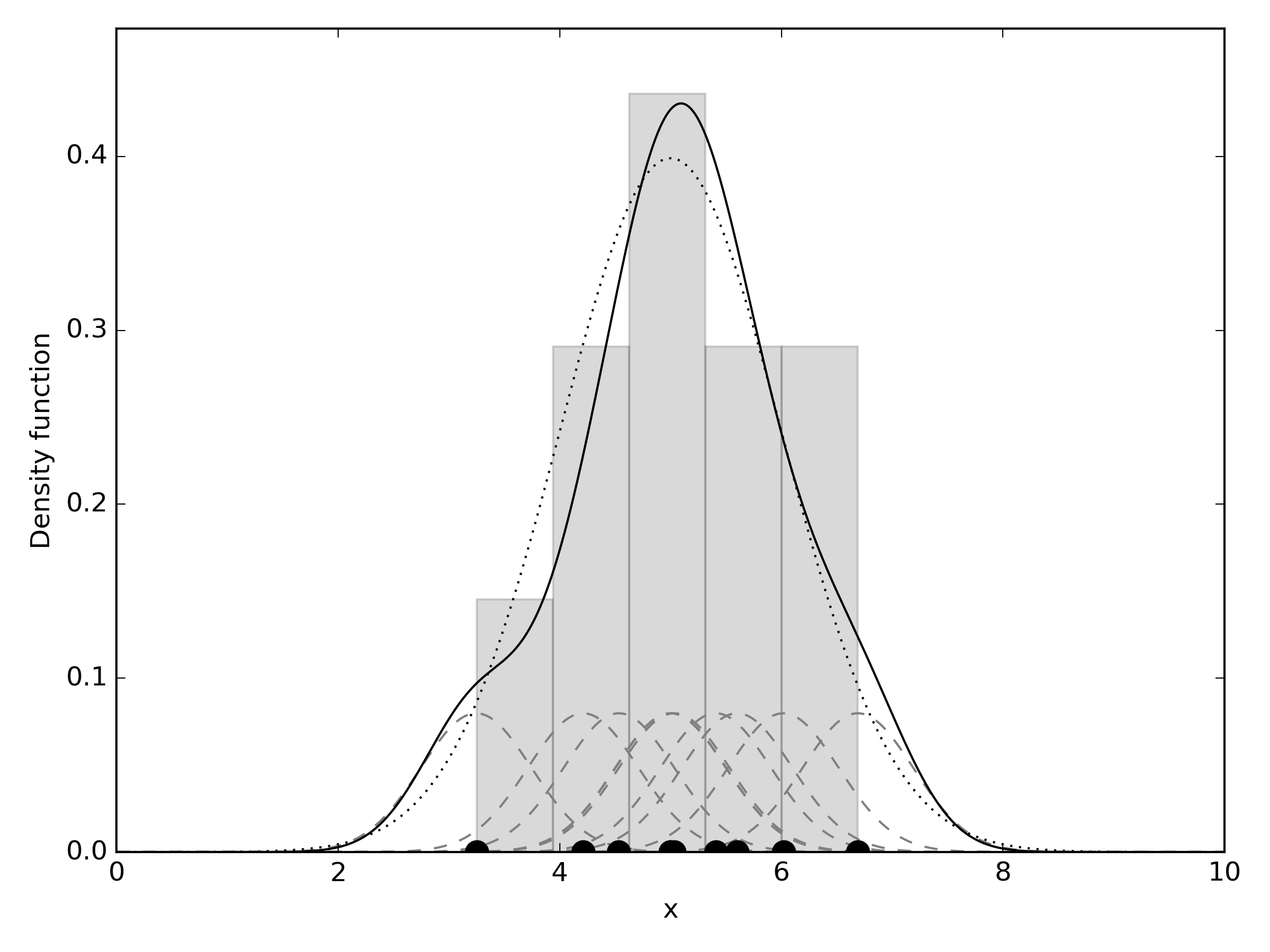}
  \caption{An example of a 1D KDE applied to 10 normally distributed points. Black dots show individual data points, while the small dashed Gaussians show individual contributions of each point to the final density estimate. The black curve represents the KDE, while a 5-bin histogram and the original normal distribution (dotted curve) are shown for comparison in the background.}
  \label{fig:1d_kde}
\end{figure}

Figure \ref{fig:1d_kde} gives an example of a Gaussian kernel density estimation applied to 10 points drawn from the normal distribution $\mathcal{N}(5, 1)$: 6.69, 4.53, 5.03, 5.41, 4.21, 5.0, 5.0, 3.25, 6.02, and 5.6. The KDE was run with a bandwidth of 0.5. The black curve shows the KDE estimate, while a 5-bin histogram and the original normal distribution (dotted) are shown in the background for comparison. The KDE distribution peaks correctly at $x = 5$ and continuously drops off to both sides. Comparing the KDE to the histogram in the background, it can be seen that the histogram overestimated the density at the $6 < x < 6.8$ interval by over 50\%, while the KDE estimation is closer to the original distribution. Furthermore, the KDE properly estimates the tails of the distribution, which the histogram completely ignores due to a small number of samples. Each point effectively serves as a ``vote" in the observed parameter space, and the shape of the ``vote" is defined by the kernel. The KDE is the sum of all votes taking also into consideration their respective shapes. At this point it is important to notice the correlation between the number of data points and the bandwidth - the more data we have, the lower the desired bandwidth. Choosing too large a bandwidth leads to overestimation of the amplitude of the true underlying distribution. The bandwidth is acting as a filter - the larger the bandwidth the lower the equivalent bandpass of the filter and vice versa. 

The KDE can be extended to an arbitrary number of dimensions, $d$ \citep{simonoff1996smoothing}:

\begin{equation}
\hat{f}(\vec{x}) = \frac{1}{n|\vec{H}|} \sum\limits_{i=1}^{n} K_d \left[ \vec{H^{-1}}(\vec{x} - \vec{x_i}) \right]
\end{equation}

where $\vec{x} = (x_1, x_2, ..., x_d)^T, \vec{x_i} = (x_{i1}, x_{i2}, ..., x_{id})^T, i=1, 2, ..., n $. $\vec{H}$ is the $d \times d$ bandwidth matrix, while the kernel $K_d$ is the appropriate multivariate kernel function.

The multivariate KDE estimates the density in each dimension with respect to all mutual correlations. This is an advantage compared to a histogram-based approach which generate parameters from each dimension separately, thus losing the underlying correlation structure in the data as we have previously noted in our examination of Method E in the introduction. The contribution of each point is smoothed out in each dimension according to the given kernel and bandwidth, giving a continuous density estimate. The disadvantage of the multivariate approach is that one also needs to estimate the covariance matrix, a non--trivial exercise.

\begin{figure}
  \includegraphics[width=\linewidth]{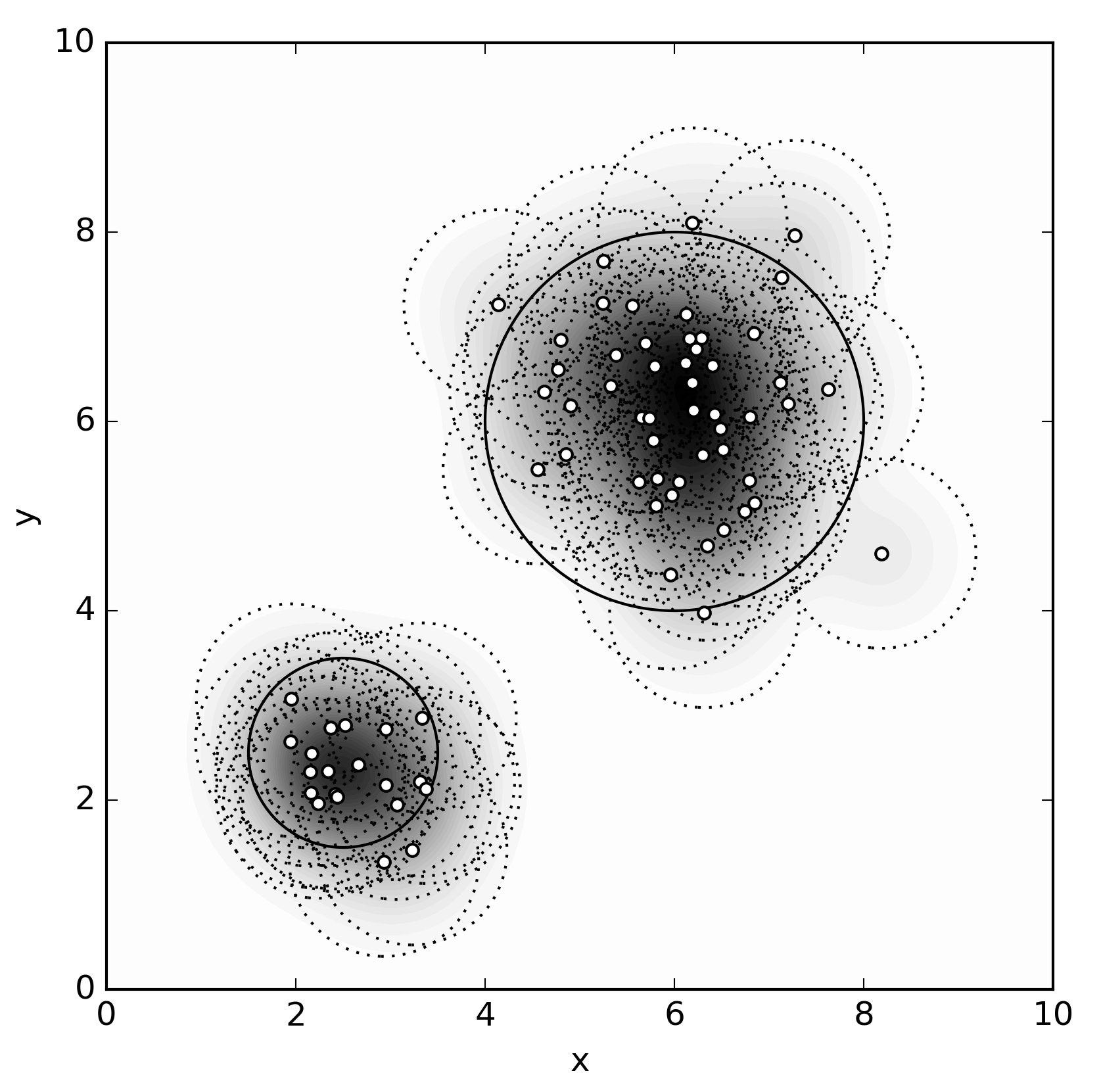}
  \caption{An example of a 2D KDE of a bimodal bivariate Gaussian distribution.}
  \label{fig:2d_kde}
\end{figure}

As a concrete example, consider a density estimate of points drawn from two bivariate normal distributions as given in Figure \ref{fig:2d_kde}. The first distribution has a mean at (2.5, 2.5) and a covariance matrix of $(0.5)^2 \vec{I_2}$, while the other has a mean at (6, 6) and a covariance matrix of $(1.0)^2 \vec{I_2}$. In all 20 points where drawn from the first, and 50 from the second distribution. The black circles show $2\sigma$ contours of each distribution. The KDE was done using a Gaussian kernel with a covariance matrix of $(0.5)^2 \vec{I_2}$ and bandwidth $h = 1.0$. Dotted circles around each point represent the $2\sigma$ contours of the kernel. Finally, the KDE is shown as the background color in the plot, darker areas representing higher density estimates. Clearly, if one draws samples out of the corresponding KDE, the picks will reflect the correlation between the dimensions and thus be a closer proxy to the original data than the samples drawn from dimensionally independent histograms.

\section{Generating instrumentally specific synthetic meteoroid orbits using the KDE}

To apply the KDE in practice, orbits were extracted from the filtered CAMS database (as outlined in Introduction), where $q_i$, $e_i$, $i_i$, $\Omega_i$, and $\omega_i$ denote the Keplerian orbital parameters of meteoroid orbit $i$, representing perihelion distance (in Astronomical Units), eccentricity, inclination, longitude of the ascending node, and argument of perihelion, respectively. Note that angular parameters are converted to radians before using them in the actual algorithm, but the results are presented in degrees for convenience.

Once these distributions from the observed dataset are obtained, the KDE input data vector is formed: 

\begin{eqnarray}
\vec{x_{i}} = \left( q_i, e_i, i_i, \Omega_i, \omega_i \right)^T
\end{eqnarray}

The number of dimensions is thus $d = 5$. Note that only 4 free orbital parameters are needed for most Earth-impacting meteoroids - the exception are low-inclination meteoroids whose nodal distances may lie outside the vicinity of Earth's orbit \citep{jopek2010remarks}. Nevertheless, as this peculiarity is naturally present in the dataset, it is organically included in the KDE analysis (together with any other instrument or site-specific biases). 

\subsection{Selecting the bandwidth}
As computed and described in \cite{jenniskens2016cams}, the database contains the standard deviation of each parameter for each orbit. The arithmetic mean standard deviation across all orbits used in the observed dataset  was calculated and found to be: $\bar{\sigma}_q = 0.0098 AU$, $\bar{\sigma}_e = 0.0337$, $\bar{\sigma}_i = 1.0267\degree$, $\bar{\sigma}_\Omega  = 0.0896\degree$, $\bar{\sigma}_\omega = 2.2756\degree$. We note that the $\bar{\sigma}_\Omega$ in the raw data is high, as $0.09\degree$ in node would nominally translate into about 2 hours in absolute time uncertainty (while the system had much more precise timekeeping). In fact, these large values are an expression of the fact that at very low inclinations it is possible to have large spreads in the ascending node which do not correspond strictly with uncertainties in time of appearance \citep{Clark2011}. While our value of $\bar{\sigma}_\Omega$ does not have a physical meaning, as an operational value it produced good results. Upon using a more realistic value of $\bar{\sigma}_\Omega = 0.001$, a mean value of nodes of orbits with inclination in the $20\degree < i < 160\degree$ range, no differences were noticed in produced results. Thus, to preserve consistency, no such special case for $\bar{\sigma}_\Omega$ was used ($\bar{\sigma}_\Omega$ was left at 0.09). Moreover, it is important to realize that these values are used only as a rough approximation of scaling between individual parameters which will later be individually scaled using empirically derived values. 

For the kernel, a 5 dimensional multivariate Gaussian kernel was used and the covariance matrix $\vec{Q}$ was assumed to be diagonal. In reality the real covariance matrix may be much more complicated, but our task is simplified to finding a diagonal covariance matrix for which the KDE operationally reproduces the bulk of the apparent correlation between elements. On the diagonal, squared mean values of standard deviations of each dimension were positioned:

\begin{eqnarray}
\vec{Q} = \begin{bmatrix}
\bar{\sigma}_q^2 &      0           &      0 &           0 & 0 \\
      0          & \bar{\sigma}_e^2 &      0 &           0 & 0 \\
      0          &      0           & \bar{\sigma}_i^2 & 0 & 0 \\
      0          &      0           &      0           & \bar{\sigma}_\Omega^2 & 0\\
      0          &      0           &      0           &   0 & \bar{\sigma}_\omega^2
\end{bmatrix}
\end{eqnarray}

This covariance matrix was used as a part of the Gaussian kernel, thus setting the final shape of ``votes" of individual orbits. It is necessary to choose an appropriate bandwidth matrix which regulates the size of the footprint of each vote. As we already have our starting covariance matrix using the known standard deviations from the data as a starting point, the bandwidth matrix scales the values in the covariance matrix to best match the observed data in a forward modelling sense. Thus, the bandwidth matrix was chosen to be a scalar matrix:

\begin{eqnarray} \label{eq:scalar_bandwidth_matrix}
\vec{H} = h \vec{I_5}
\end{eqnarray}

where the value $h$ regulates the total size of the kernel and the significance of each vote in the 5-dimensional parameter space. If one uses many data points with a high bandwidth, we expect that the resulting KDE will not preserve all the details of the original distribution. On the other hand, if the bandwidth is too small, individual contributions would be too far apart and they will not reveal the underlying structure in the data at all. In practice, this means that when drawing samples from the resulting KDE, using a small bandwidth results in synthetic data too similar to the input data points.

Our initial attempt assumed a scalar matrix for simplicity, but we found this inadequate to correctly reproduce all aspects of the data (see section \ref{sec:scalar_bandwidth_results}). Thus, an alternative (more complex) approach was employed to use a diagonal bandwidth matrix with a different bandwidth for each parameter:

\begin{eqnarray} \label{eq:nonscalar_bandwidth_matrix}
\vec{H} = \begin{bmatrix}
  h_q      &      0        &      0     &        0     & 0 \\
   0       &     h_e       &      0     &        0     & 0 \\
   0       &      0        &     h_i    &        0     & 0 \\
   0       &      0        &      0     &    h_\Omega  & 0\\
   0       &      0        &      0     &        0     & h_\omega
\end{bmatrix}
\end{eqnarray}
The results using a non-scalar bandwidth matrix are discussed in section \ref{sec:non-scalar_bandwidth_results}.

Although the Earth orbit intersection condition (Equation \ref{eq:earth_crossing}) was used to estimate the perihelion distance in \cite{jopek2016probability}, it was not utilized as a part of this method - $q$ was directly estimated from the density model. The main reason for not using this equation is the loss of correlation in the perihelion distance with other orbital elements. For example, if it is assumed that the meteoroid's heliocentric distance at the moment of collision with Earth is fixed at 1 AU at all positions in the orbit, the real range of orbits is not covered, as the Earth's distance to the Sun slightly changes with time. Moreover, \cite{jopek2016probability} method E uniformly distributes the $e \cos{\omega}$ signs, not taking into account any correlations in the orbits other than the ecliptic latitude ratio. This option was tested and it was found that more consistent results were obtained when $q$ was retained as one of the estimated parameters. On the other hand, equation \ref{eq:a_calc} was still used to calculate the semi-major axis, as it is a simple geometrical relation with no additional assumptions.

\subsection{Sampling the KDE} \label{subsec:kde_sampling}

To draw samples out the KDE, the following algorithm was applied:
\begin{enumerate}
    \item A random point $\vec{x_i}$ was drawn out of the set of input points $\vec{x_1}, ... \vec{x_n}$.
    \item A sample point is drawn from the multivariate Gaussian $\mathcal{N} \left( \vec{x_{i}}, \vec{H^{\circ\frac12}} \right)$ centered at $\vec{x_i}$ with the standard deviation $\vec{H^{\circ\frac12}}$ (Hadamard root, i.e. the element-wise square root of the bandwidth matrix).
    \item Repeat until the desired number of samples is drawn.
\end{enumerate}    

The Mersenne Twister algorithm was used to generate random numbers (Python numpy library implementation, see section \ref{implementation details} for more details).
As the analysis did not accommodate circular parameters, some of the angular orbital elements were generated out of the $[0, 2\pi]$ range. For those cases, the values of $\omega$ and $\Omega$ outside the $[0, 2\pi]$ range were wrapped inside the allowable range, while the values of $i$ outside the $[0, \pi]$ range were mirrored with respect to $0$ for values of $i < 0$, and with respect to $\pi$ for values of $i > \pi$.
A culling of orbits in the synthetic distribution which were not in the initial range of observed orbital parameters was performed, for each parameter. Orbits with $1/a$ outside the $[0.0010, 1.8587]$ range, $q$ outside the $[0.0013, 1.0167]$ range, $e$ outside the $[0.0120, 1.0000]$ range were rejected and new orbits were drawn until the total number of synthetic orbits matched the number of observed orbits. This ensured that the data is indeed inside the initial limits. It is worth pointing out that this procedure does not undermine the KDE's effectiveness in estimating tails of distributions, as these boundaries are indeed physical. If we were to allow eccentricities larger then 1, an artificial bias towards a larger number of extrasolar meteoroid would be introduced. The perihelion distance boundary is physical as well, as it cannot extend beyond the Earth's maximum distance from the Sun. The estimate of numbers of orbits at the local level, i.e. tails of distributions estimating individual sporadic sources, are preserved.

\section{Scalar bandwidth matrix results} \label{sec:scalar_bandwidth_results}
In this section, we explore how changes in the scalar bandwidth matrix (equation \ref{eq:scalar_bandwidth_matrix}) affect final results.
As the goal of our effort is to preserve the underlying structure and statistics of the observed data, so that each model run is representative of the sporadic background as measured by a particular instrument, a valid approximation for the bandwidth $h$ is needed. For the structural comparison, it was decided to compare 2D histograms of the observed and synthetic data created with a range of bandwidth values. The bin sizes for each dimension in the histogram were chosen according to the standard deviation of each orbital element. For each bandwidth examined, the same number of synthetic orbits as input orbits was drawn from the KDE. We focus on the $\omega$ vs. $q$ histogram, as these parameters show the strongest visible dependency (see Figure \ref{fig:peri-q_jopek}), which we wish to preserve. Furthermore, as the bandwidth value scales all parameters linearly, it was sufficient to take only one type of histogram into consideration for testing, though in the final analysis we check all combinations. 

\begin{figure}
  \includegraphics[width=\linewidth]{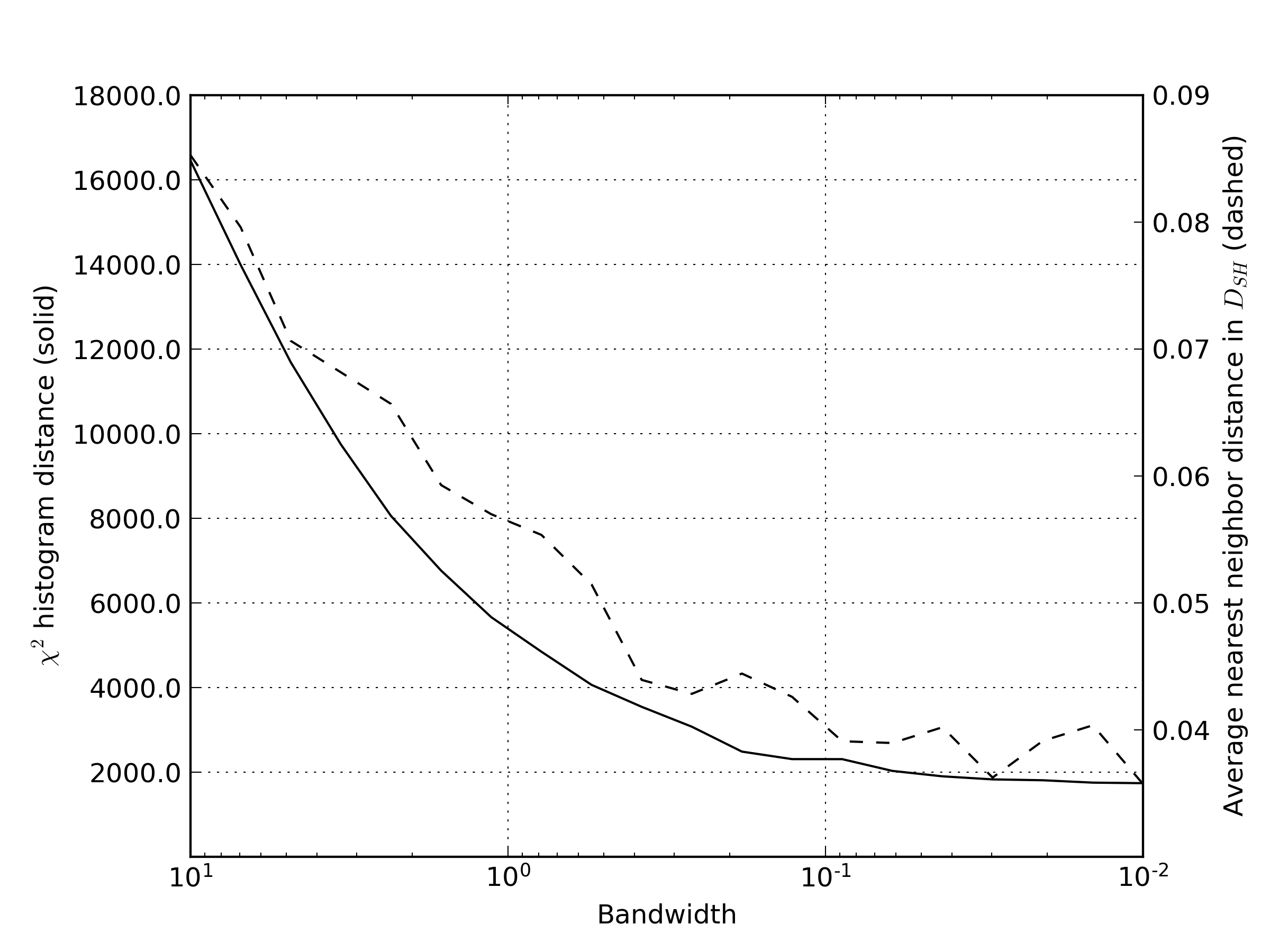}
  \caption{Results of a comparison of $\omega$ vs. $q$ for synthetic orbits vs. observed orbits (solid line) and the mean nearest neighbour distance (dashed line), for bandwidth values from 10 to 0.01.}
  \label{fig:bandwidth_analysis}
\end{figure}

The number of bins for each parameter was chosen on the basis of the mean standard deviations: the argument of perihelion was divided into $k_{\omega} = 360/\bar{\sigma}_\omega = 158$ bins, while the perihelion distance was divided into $k_{q} = 1.0167/\bar{\sigma}_q = 104$ bins, thus making a 2D histogram with $158 \times 104$ bins. To assess the statistical similarity between histograms, the $\chi^2$ histogram distance \citep{pele2010quadratic} was used:

\begin{eqnarray} \label{eq:chi_sq_hist_dist}
\chi^2 \left(P, Q \right) = \frac{1}{2} \sum\limits_{i=1}^{k_{\omega}} \sum\limits_{j=1}^{k_{q}} \frac{\left(P_{ij} - Q_{ij} \right)^2} {\left(P_{ij} + Q_{ij} \right)}
\end{eqnarray}

where $P_{ij}$ are bins of synthetic data histogram, while $Q_{ij}$ are bins of observed data histogram. For the bins with no counts, it was assumed that $\frac{0}{0} = 0$. This procedure was repeated 10 times for each bandwidth value and results of iterations were averaged to avoid the influence of randomness during the creation of synthetic orbits. To estimate the statistics and the density of the generated data, the arithmetic mean of the nearest neighbour (hereinafter referred to as NN) distance (i.e. the minimum distance from each point to any other, except itself) was calculated using the \cite{southworth1963statistics} D criterion. These two parameters were calculated for a range of bandwidth values from 10 to 0.01. The results are shown in Figure \ref{fig:bandwidth_analysis}. It can be seen from the figure that the histogram distance (solid line) sharply decreases until we reach a bandwidth value of 0.1, meaning that the differences in orbits being tested has gone below the resolution of the histogram. As the histogram was constrained by the original standard deviation of the observed data, this serves to effectively determine the bandwidth at which the observed and the synthetic data can no longer be visually differentiated. The mean NN distance (dashed line) follows the same trend, falling sharply until we reach a bandwidth of $h = 0.1$. For comparison, the original (observed) dataset had a mean NN distance in $D_{SH}$ of 0.08, while the value for the synthetic orbits obtained by the \cite{jopek2016probability} method was 0.1014. If the NN distance is chosen to match the observed data at $h = 10$, the histogram distance is then high. On the other hand, if we choose to preserve the structure and also choose a lower value of histogram distance at $h = 0.1$, the mean NN distance is half that of the observed data. As we desire to preserve both the original structure (i.e. the lower histogram distance) and the underlying statistics (i.e. the same NN distance as the original), we conclude that this method requires a trade-off: either the structure in the data is preserved or the statistics, but not both.

The high values of bandwidth are not preserving the structure because the contribution of each point of original data to the KDE parameter space is spread our over a large ``area" (the explanation uses a 2D Gaussian as an example, but in reality the contributions are spread out in 5 dimensions), but the amplitude is small as the volume of the Gaussian is unity at all bandwidths. This makes the resulting KDE to overestimate the density of sparsely populated parts of the parameters space, and underestimate the density of parts with abundant number of points in the original distribution. For a low bandwidth value, the opposite is true. 

To further explore the effect of bandwidth differences, the analysis was repeated for two extreme values of the bandwidth, $h = 10$ and $h = 0.1$. Moreover, a problem might be that the shape of the kernel which is used for ``voting" in the parameter space is not ideal, which can cause an issue where no optimal bandwidth can be found, as the density of one dimension will always be underestimated, while the other will be overestimated. As the D criterion, used as our statistical measure, reduces all dimensions to one single value, this asymmetry in densities per different parameters can influence the estimate of mean distances between individual data points as well. To counter this issue, we are discussing the usage of a non-scalar bandwidth matrix in section 5.

\subsection{Results with bandwidth h=10}
The synthetic orbits obtained using $h = 10$ were analysed in the same manner as in Section 1, using histograms and density plots. These data had the same mean NN distance as the original dataset, but show larger differences in the $\omega$ vs. $q$ histogram.

\begin{figure}
  \includegraphics[width=\linewidth]{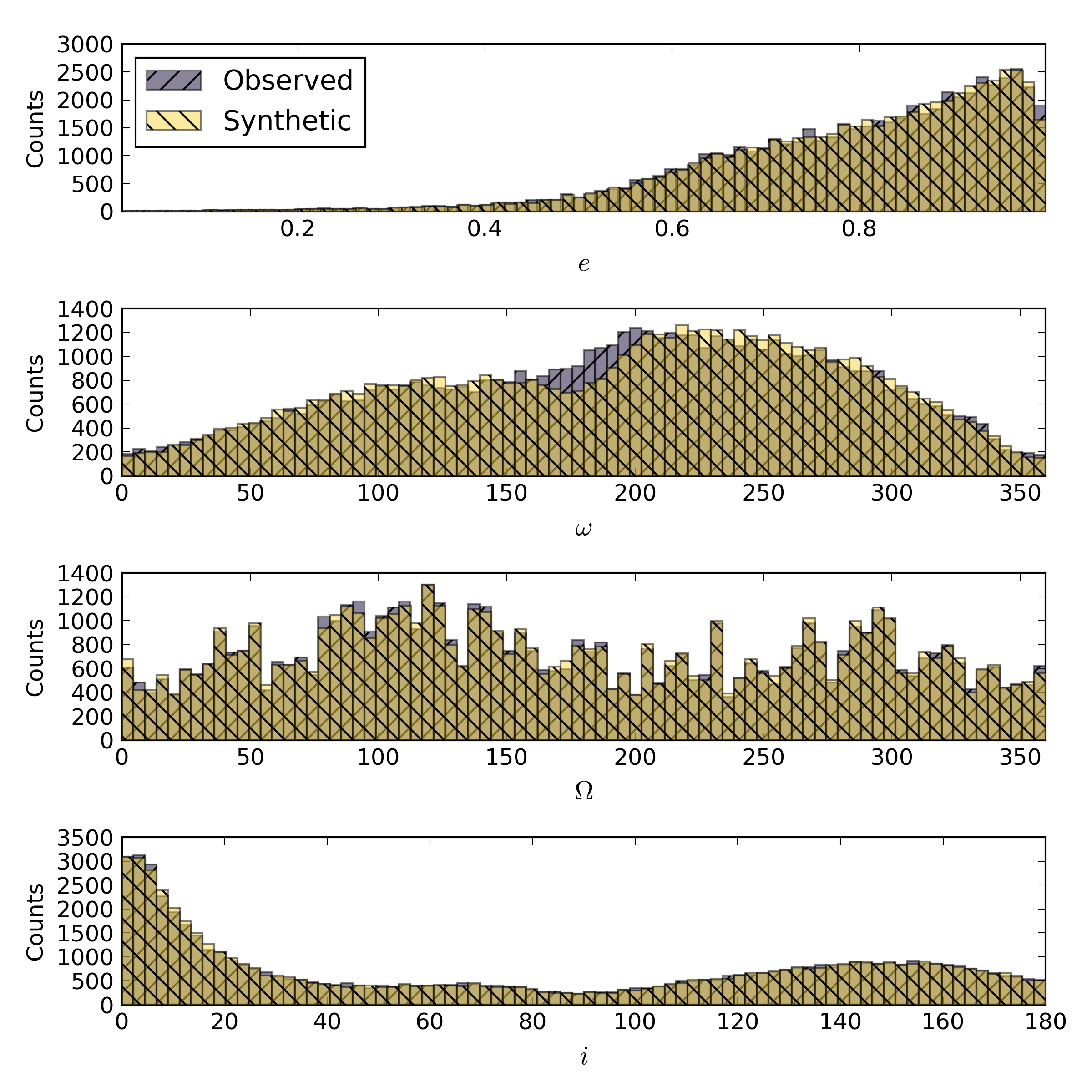}
  \caption{Comparison of histograms of all generated orbital parameters: $e$, $\omega$, $\Omega$ and $i$, using the newly proposed method and $h=10$.}
  \label{fig:bandwidth_10_all}
\end{figure}

Figure \ref{fig:bandwidth_10_all} shows a comparison of observed and synthetic orbit histograms per individual orbital parameters. The figure shows that the differences are minor, except an apparent dearth of orbits around $\omega \approx 180\degree$. Otherwise, the results are similar to those obtained by the \cite{jopek2016probability} method.

\begin{figure}
  \includegraphics[width=\linewidth]{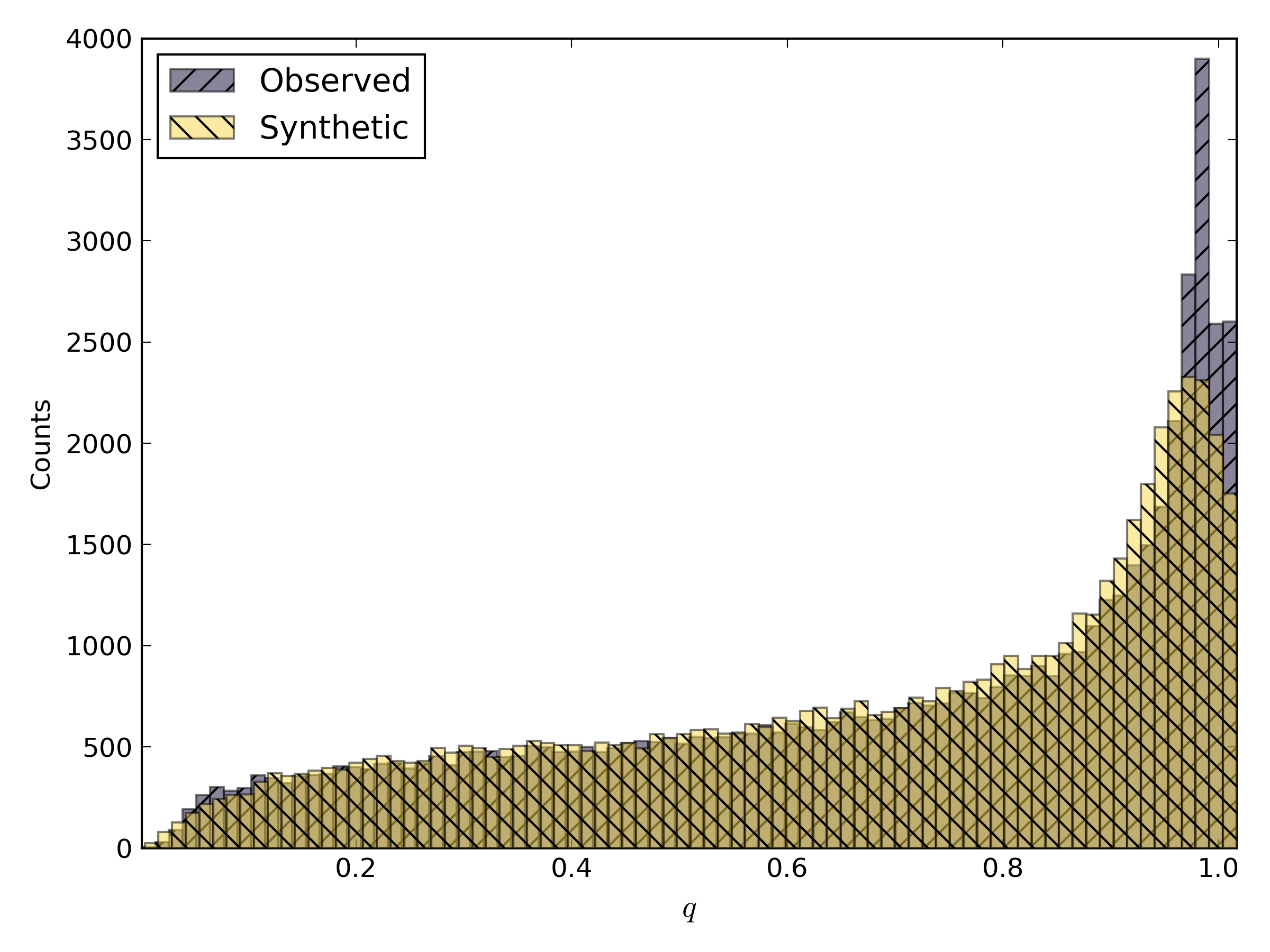}
  \caption{Perihelion distance histogram produced by the KDE method and $h = 10$.}
  \label{fig:bandwidth_10_q}
\end{figure}

The difference between methods starts to show in the perihelion distance histogram, shown in Figure \ref{fig:bandwidth_10_q}. In this case the lower values of $q$ are correctly reproduced; however there is an apparent shortage of orbits at $q \approx 1$, where about 1500 bin counts are missing, and are shifted towards lower values of $q$. The original range of the data has been correctly reproduced, the maximum value of $q$ goes just beyond 1 AU, although the number of counts differ.

\begin{figure}
  \includegraphics[width=\linewidth]{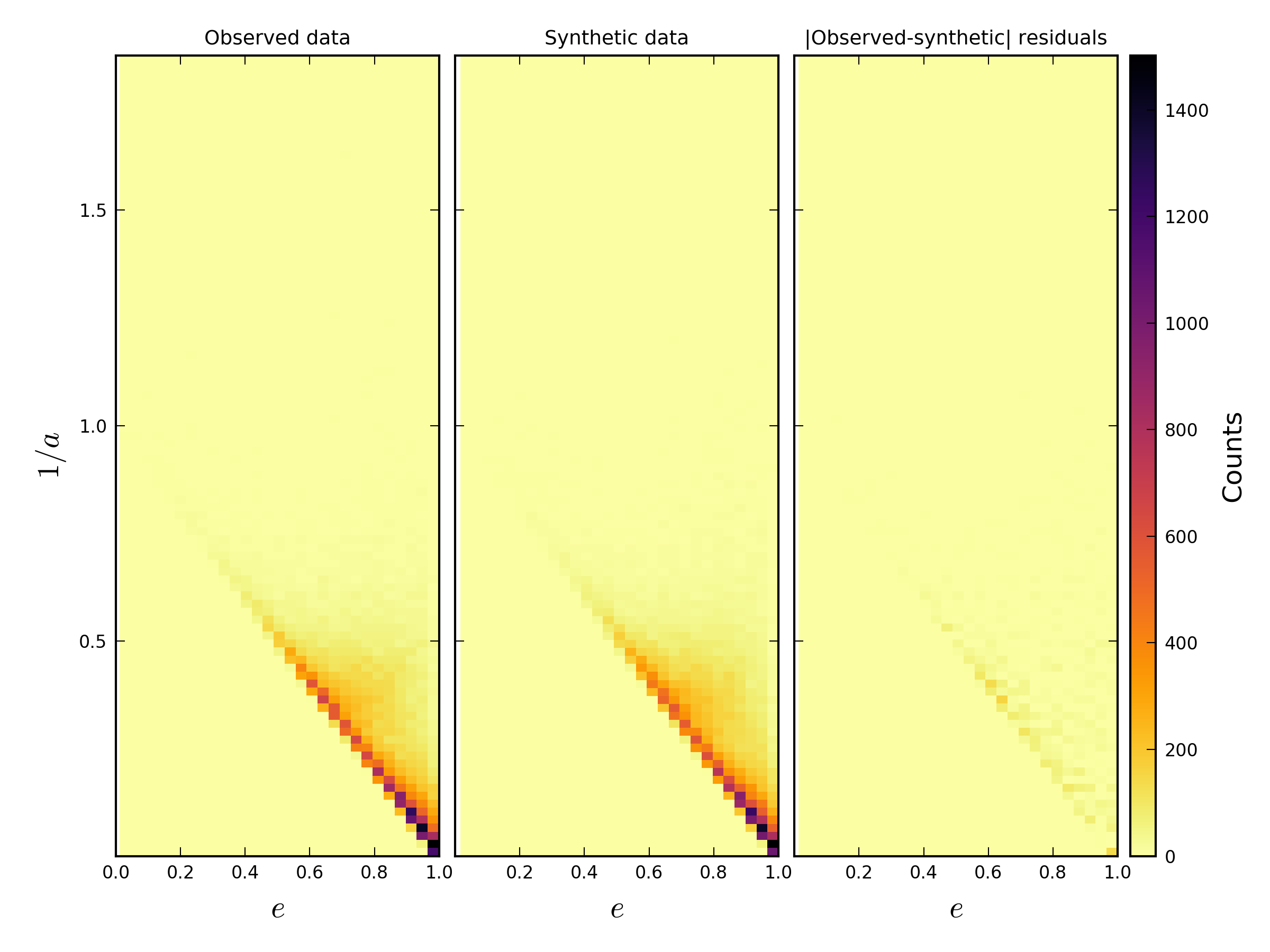}
  \caption{Eccentricity vs. inverse semi-major axis 2D histograms produced by the KDE method using $h = 10$.}
  \label{fig:bandwidth_10_e_vs_inv_a}
\end{figure}

When comparing the $e$ vs. $1/a$ density plots obtained by the new method, shown in Figure \ref{fig:bandwidth_10_e_vs_inv_a}, it can be seen that the general trend in the data was correctly reproduced, as well the area of higher density at $e \approx 0.6$. Furthermore, the diffuse area of lower density above the main trend line has also been reproduced; the overall residuals are low.

\begin{figure}
  \includegraphics[width=\linewidth]{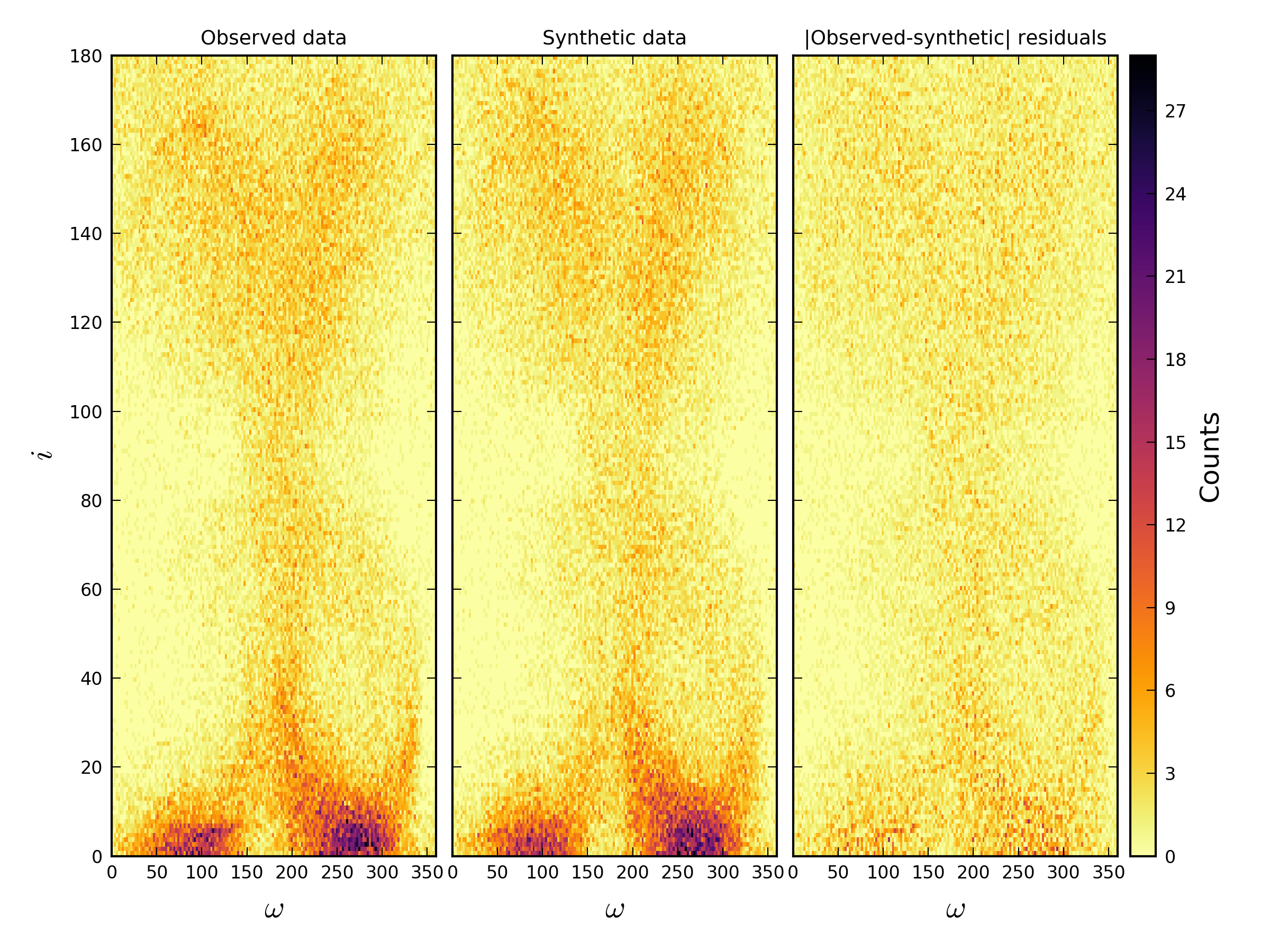}
  \caption{Argument of perihelion vs. inclination 2D histograms produced by the KDE method and $h = 10$.}
  \label{fig:bandwidth_10_peri_vs_i}
\end{figure}

Figure \ref{fig:bandwidth_10_peri_vs_i} shows a comparison of density plots of $\omega$ and $i$. The main concentrations of orbits have been properly reproduced, thus the residuals plot is uniform, although there are several randomly distributed bins showing high residuals. An area of higher density at $\omega \approx 260\degree$ and low inclinations is not as populated as in the observed data, thus the high residual counts in that area. However, comparing the results to Figure \ref{fig:incl-peri_jopek} where the results of the \cite{jopek2016probability} method was shown, a significant improvement in residuals is apparent.

\begin{figure}
  \includegraphics[width=\linewidth]{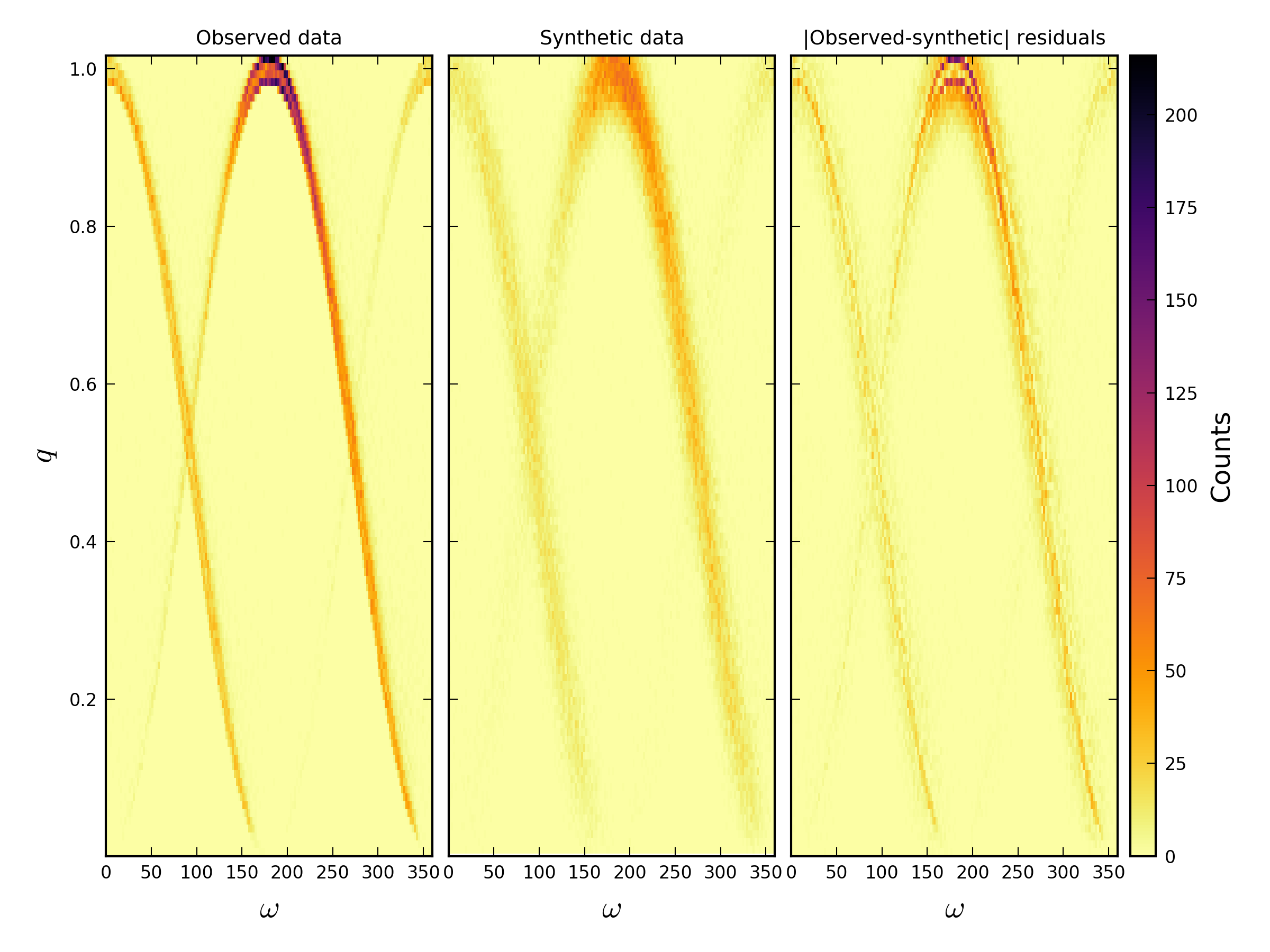}
  \caption{Argument of perihelion vs. perihelion distance produced by the KDE method using $h = 10$.}
  \label{fig:bandwidth_10_peri_vs_q}
\end{figure}

Finally, Figure \ref{fig:bandwidth_10_peri_vs_q} shows density plots of $\omega$ vs. $q$. The main features of the observed dataset have been reproduced, though the synthetic data plot appears to be blurred compared to the original. Nevertheless, the general density distribution has been preserved and both main branches are visible. This is not the case in the \cite{jopek2016probability} synthetic data. On the other hand, the low and high $q$ central branches were not reproduced as shown in the residuals inset. This characteristic of the synthetic data is the only one that significantly differs from the observed dataset. The underlying reason is apparent when looking at the individual histograms for $\omega$ and $q$, as these indeed differ from the original dataset. The bandwidth analysis shown in Figure \ref{fig:bandwidth_analysis} indicates that the differences in this histogram are comparatively high. On the other hand, the synthetic data generated in this manner preserves the original statistics of the 0.08 NN distance in $D_{SH}$. One possible explanation for the difference may be a bad estimation of $\bar{\sigma}_q^2$. This would explain why both the $q$ and $\omega$ vs. $q$ plots show a significant difference compared to the observed dataset. It is also possible that the estimation of $\bar{\sigma}_{\omega}^2$ could play a role, as $\omega$ participates as one of the dimensions in Figure \ref{fig:bandwidth_10_peri_vs_q}. These possibilities are explored later in the paper.

\subsection{Results with h=0.1}
\begin{figure}
  \includegraphics[width=\linewidth]{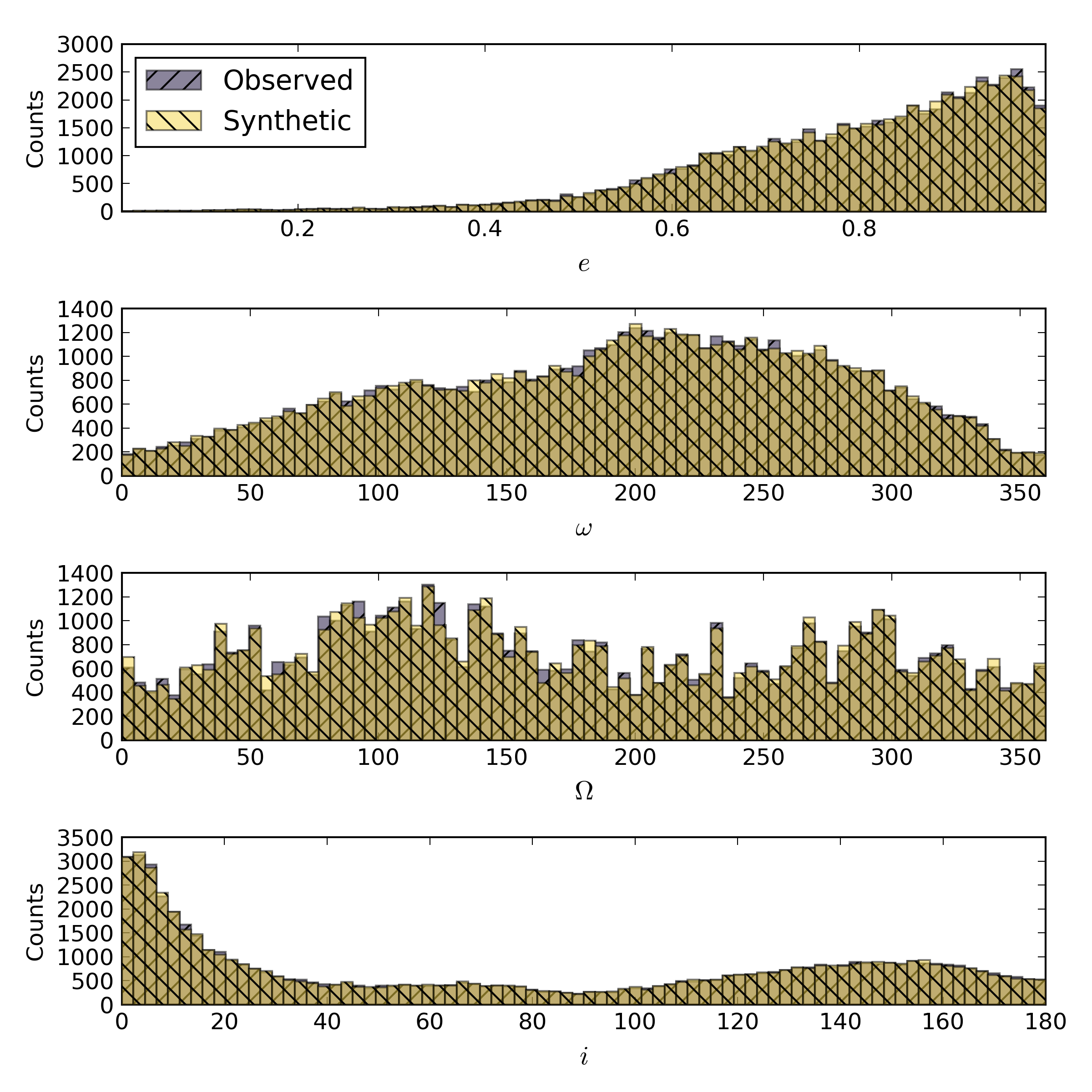}
  \caption{Comparison of observed and synthetically generated orbital parameters: $e$, $\omega$, $\Omega$ and $i$, using the newly proposed method and $h=0.1$.}
  \label{fig:bandwidth_0.1_all}
\end{figure}

Using a bandwidth $h=0.1$ produces a higher degree of similarity between the $\omega$ vs. $q$ histograms of the observed and the synthetic data compared to $h=10$, but at the expense of the data statistics as the mean NN distance in $D_{SH}$ of the synthetic data becomes 0.04 - half the value obtained for the observed data (0.08).
Figure \ref{fig:bandwidth_0.1_all} compares observed and synthetic orbit histograms  and shows that the differences are minor, the largest being about 200 counts in two bins of $\Omega$. Otherwise, the results are similar to those obtained with the \cite{jopek2016probability} method.

\begin{figure}
  \includegraphics[width=\linewidth]{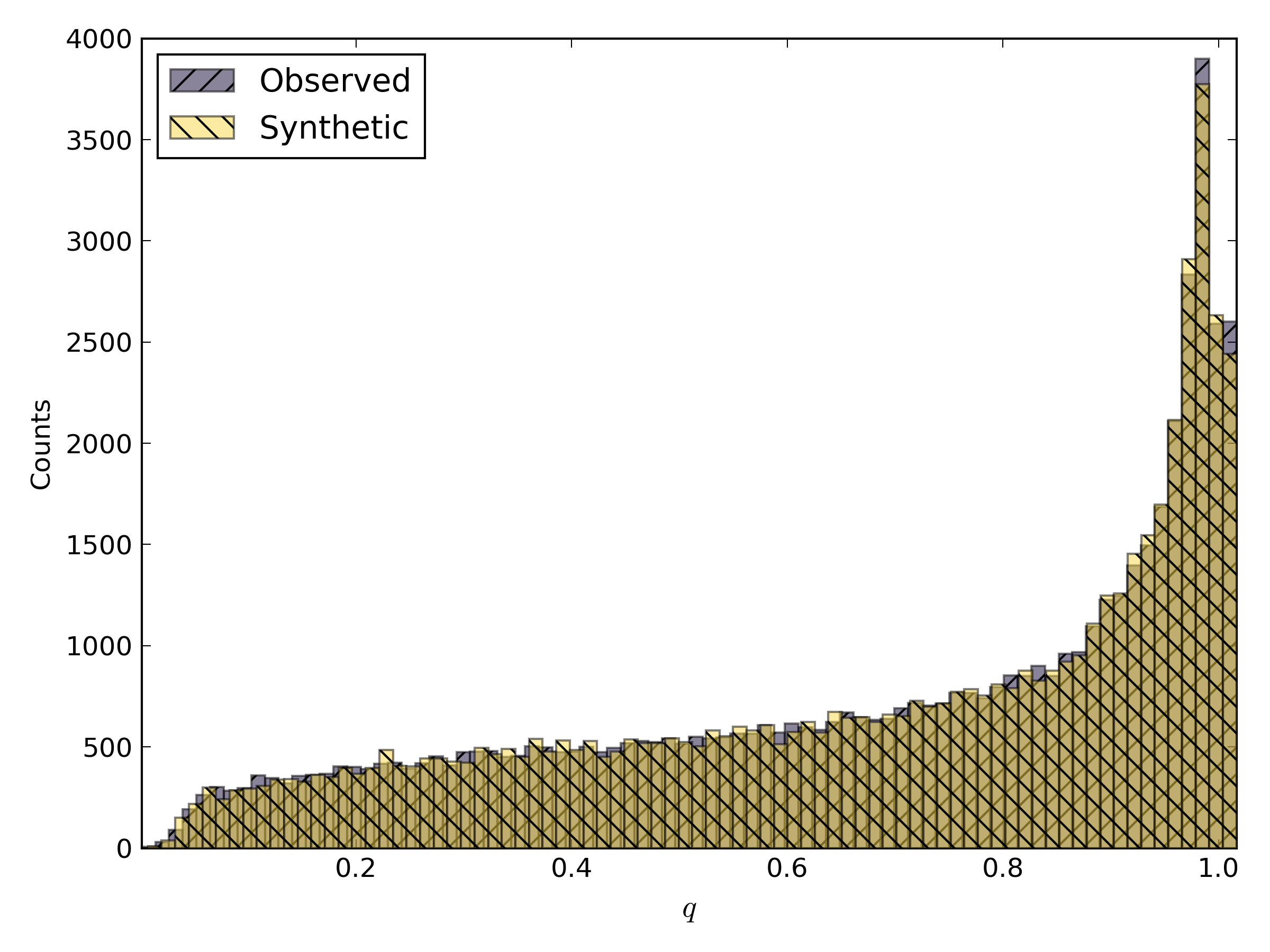}
  \caption{Perihelion distance histogram produced by the KDE method and $h = 0.1$.}
  \label{fig:bandwidth_0.1_q}
\end{figure}

We see that the difference in bandwidth mostly influences the perihelion distance histogram, shown in Figure \ref{fig:bandwidth_0.1_q}. Using $h=0.1$ gives a better reproduction of the observed histogram compared to $h=10$, with no major discrepancies in the general trend nor in the number of bin counts. The original range of the data has also been correctly reproduced, the maximum value of $q$ being just beyond 1 AU.

\begin{figure}
  \includegraphics[width=\linewidth]{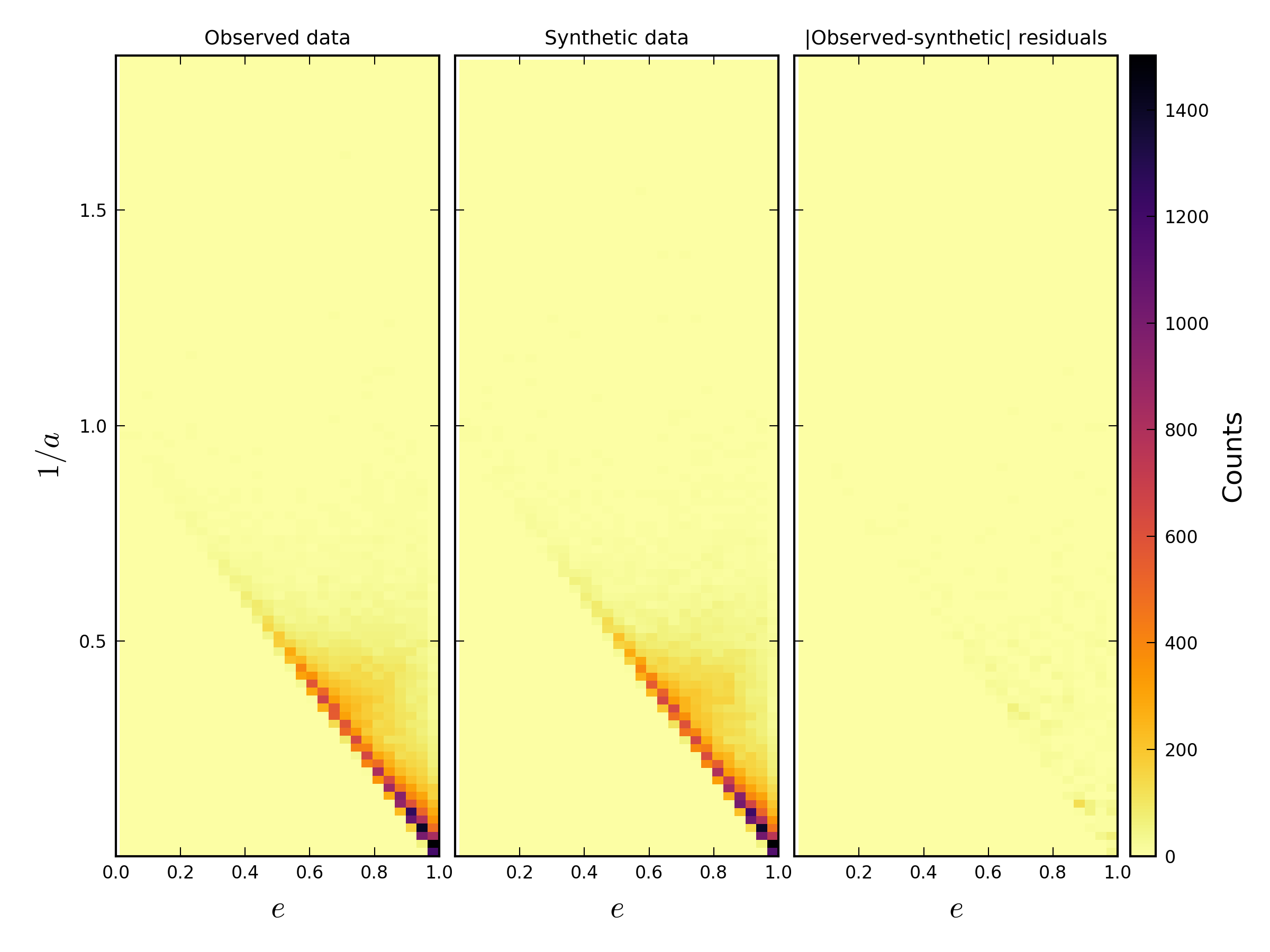}
  \caption{Eccentricity vs. inverse semi-major axis for synthetic data produced by the KDE method and $h = 0.1$.}
  \label{fig:bandwidth_0.1_e_vs_inv_a}
\end{figure}

Differences in the $h=0.1$ $e$ vs. $1/a$ density plot (Figure \ref{fig:bandwidth_0.1_e_vs_inv_a}) when compared to the $h = 10$  (Figure \ref{fig:bandwidth_10_e_vs_inv_a}) show that the lower bandwidth produced lower residuals, as expected. Nevertheless, the change in residuals is minor. This implies that the higher bandwidth value almost correctly reproduced the data range and distribution for $e$ and $1/a$;  there was no further need to reduce the bandwidth to match these parameters to the observed dataset. This behaviour suggests that our initial assumption in constructing the covariance matrix might be wrong - i.e. the standard deviations of the data do not approximate the covariance matrix correctly, or the standard deviations were not properly scaled.

\begin{figure}
  \includegraphics[width=\linewidth]{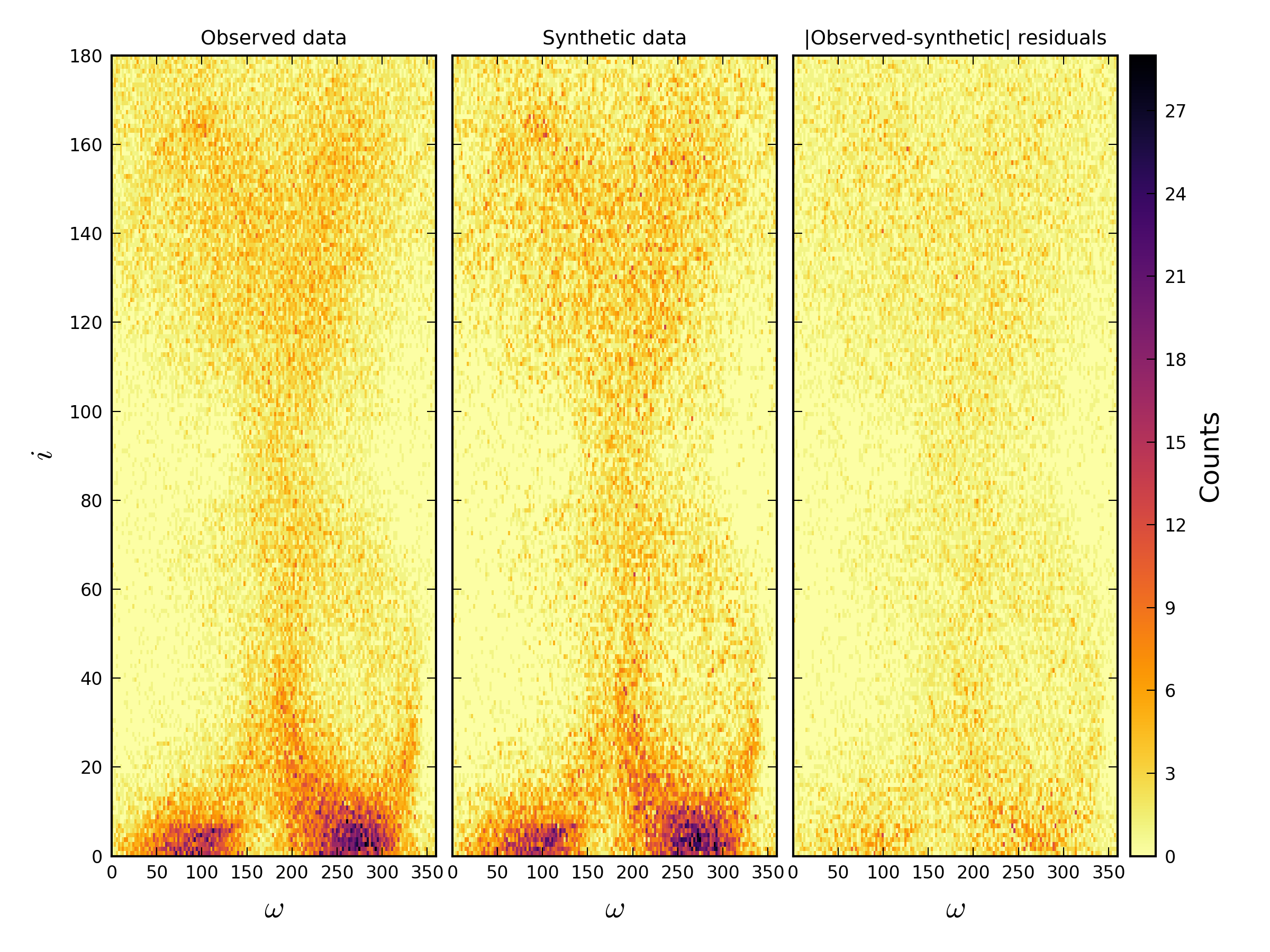}
  \caption{Argument of perihelion vs. inclination for synthetic data produced by the KDE method and $h = 0.1$.}
  \label{fig:bandwidth_0.1_peri_vs_i}
\end{figure}

This reasoning is further supported by Figure \ref{fig:bandwidth_0.1_peri_vs_i}, where $\omega$ vs. $i$ density plot is shown for $h=0.1$. The changes compared to the $h=10$ plot in Figure \ref{fig:bandwidth_10_peri_vs_i} are minor, the original structure is preserved. Nevertheless, the new graph does show certain differences: the area of the plot at $\omega \approx 270\degree$ and low inclinations is denser than on the $h=10$ plot, and this time the area at $\omega \approx 180\degree$ and high inclinations is well reproduced.

\begin{figure}
  \includegraphics[width=\linewidth]{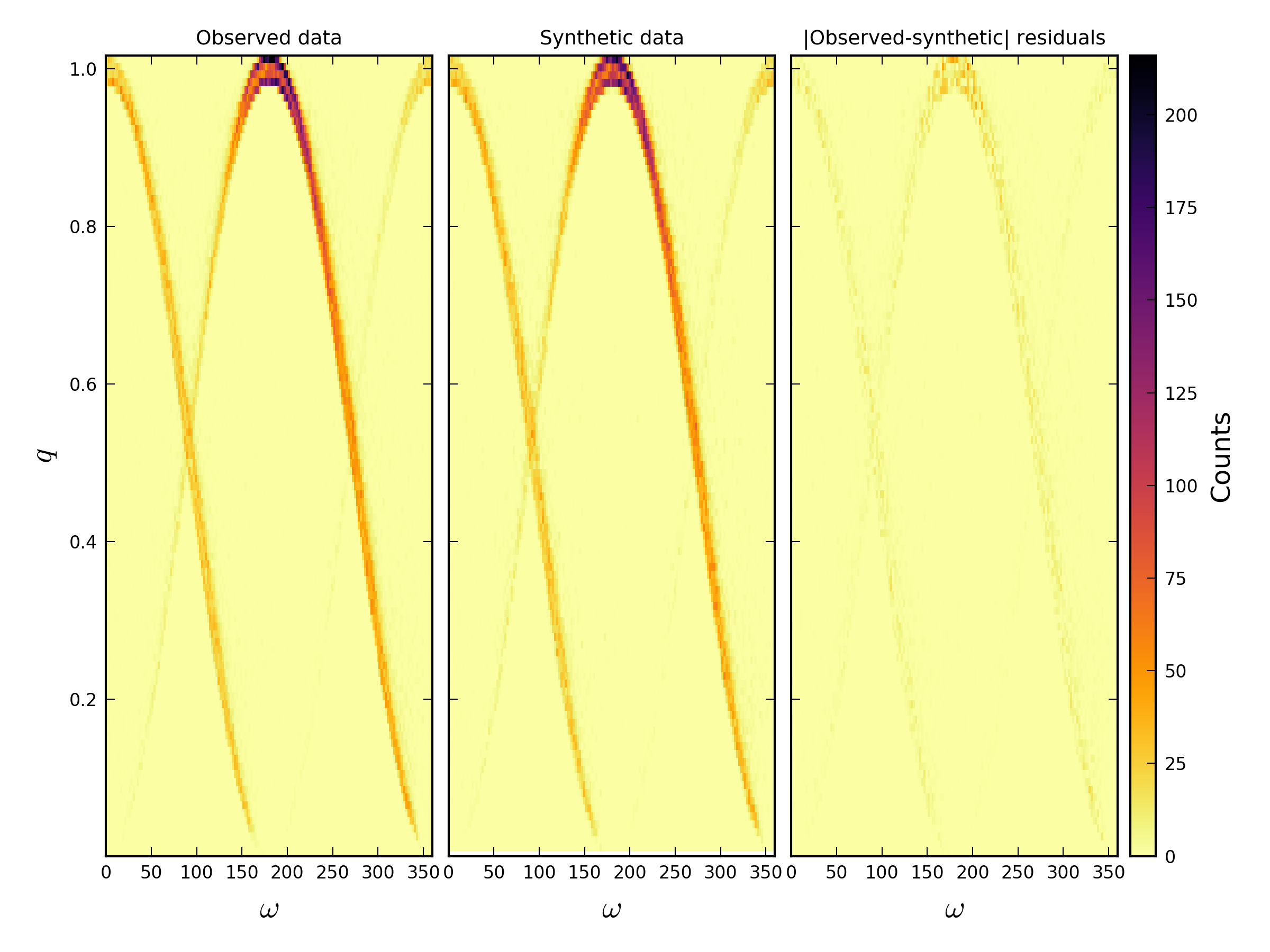}
  \caption{Argument of perihelion vs. perihelion distance for synthetic data produced by the KDE method and $h = 0.1$.}
  \label{fig:bandwidth_0.1_peri_vs_q}
\end{figure}

Finally, Figure \ref{fig:bandwidth_0.1_peri_vs_q} shows the greatest difference between using $h=0.1$ and $h=10$ (Figure \ref{fig:bandwidth_10_peri_vs_q}). For these plots with $h=0.1$, there is a much higher level of similarity between the observed and the synthetic data compared to $h=10$. The $\chi^2$ distance between the histograms is the smallest possible, as seen in Figure \ref{fig:bandwidth_analysis}. Both of the main branches are reproduced, as well as the division of the central branch into two filaments. Since the only change here is using a smaller bandwidth than before and the other orbital element correlations are relatively unchanged between $h=10$ and $h=0.1$, this leads to the conclusion that the standard deviations and a scalar bandwidth matrix did not approximate the covariance matrix well.

The graphs shown in this section reproduce the observed orbital distribution to a high degree of similarity but do not preserve the underlying statistics of the observed data. To fully understand why that is so, let us presume an ideal case where we would have a bandwidth matrix which, when used for KDE, would produce a model probability density function (PDF) identical to the real underlying PDF. If one were to draw samples out of that ideal KDE, one would end up with a set of synthetic orbits with exactly the same mean NN $D_{SH}$ as that of the observed data. In reality, we only have a rough approximation of the real bandwidth matrix, if one can even assume there exists a globally applicable bandwidth matrix for this problem. By making this assumption, in certain areas of the orbital phase space, orbits might be generated which do not correlate to each other in the same way as the observed orbits did, thus driving the global value of the mean NN $D_{SH}$ in a particular direction. As observed, for lower bandwidths, the mean NN distance is lower as well; we believe the denser areas of the orbital phase space, where the mean NN $D_{SH}$ is naturally smaller, were pushed to even smaller values due to the non-perfect bandwidth matrix. Furthermore, as a certain percentage of synthetic orbits (depending on the bandwidth) was generated outside the range of values of observed parameters, it is certain that some of the points were used more than once as the central point of a kernel (see section \ref{subsec:kde_sampling}), thus further lowering the value of our statistical indicator.

To simplify the explanation to only one dimension, let us imagine that the bandwidth for generating the KDE PDF in Figure \ref{fig:1d_kde} was smaller. Individual kernels would have higher amplitudes, but they would in turn be narrower. The points in the centre, where they are denser, would thus create a probability density estimate at the centre with a higher amplitude than that of the underlying model. If one were to sample such a KDE, more points would be drawn from the centre of the distribution, making the synthetic distribution statistically denser.

Particularly, what happened here is that due to the bandwidth being small, the shape of each ``vote" in the parameter space was narrow. This has caused areas of the parameter space which were dense in the observed data to have their density overestimated by the KDE, as the amplitude of the sum of individual local groups of Gaussians was higher than the one in the real underlying observed distribution. As a result, we decided to explore the effects of using a non-scalar bandwidth matrix.

\section{Results using a non-scalar bandwidth matrix} \label{sec:non-scalar_bandwidth_results}
From the previous sections, we find that the assumed scalar bandwidth matrix does not properly reproduce both the correlation and the statistics of the original dataset simultaneously. One possible way to remedy this shortcoming is to use  a non-scalar bandwidth matrix. Its form is given in equation \ref{eq:nonscalar_bandwidth_matrix}. As shown in the previous section, the parameters that we find most influence the final results are the perihelion distance $q$ and the argument of perihelion $\omega$. 

\begin{figure}
  \includegraphics[width=\linewidth]{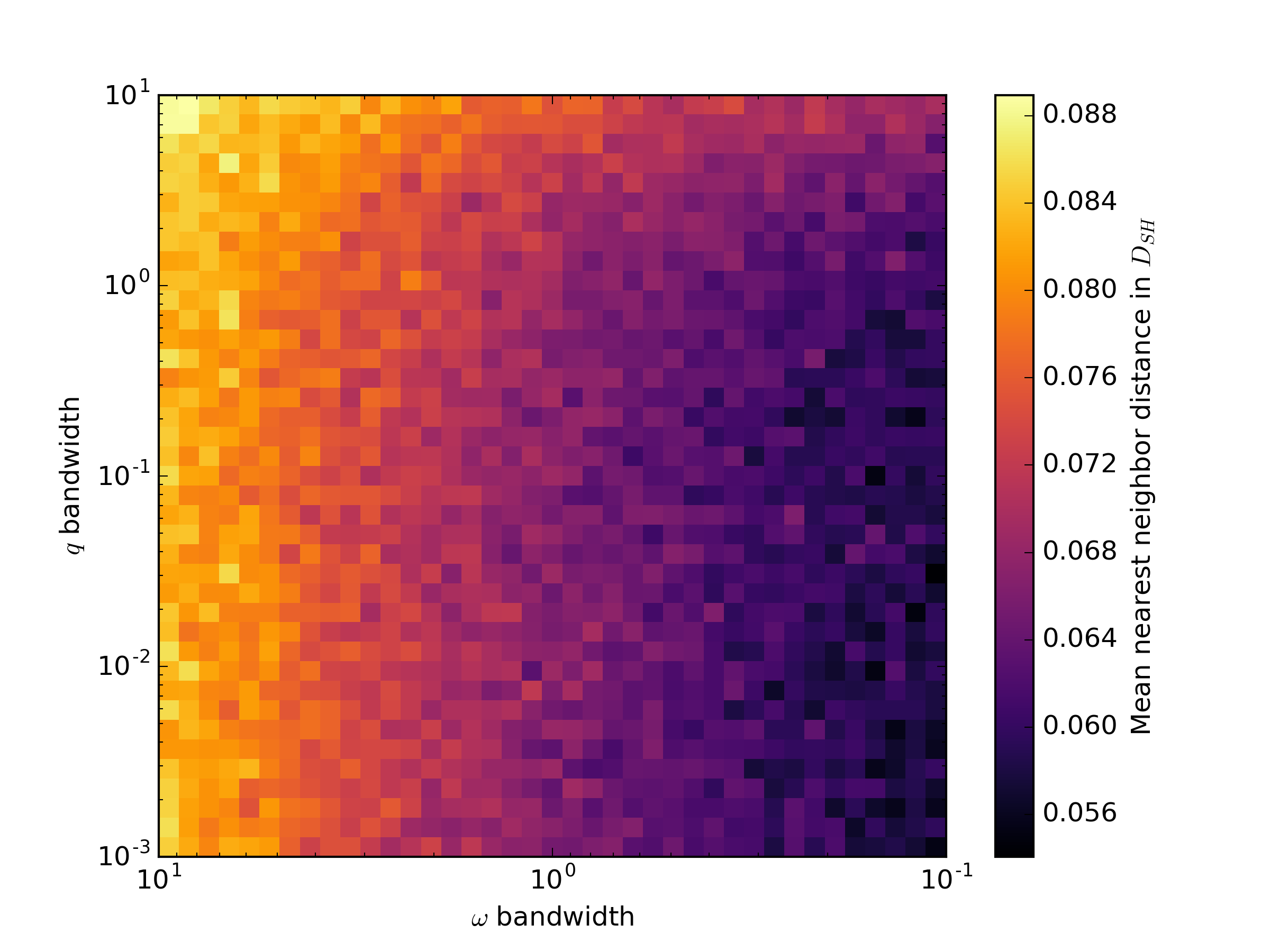}
  \caption{Influence of different values of $h_q$ and $h_\omega$ on the synthetic data statistics.}
  \label{fig:bandwidths_matrix_peri_q_nn}
\end{figure}

To examine the influence of these parameters on the data statistics we kept all  bandwidth values fixed at $h_e = h_i = h_\Omega = 10$. The $h_q$ and $h_\omega$ values were then varied between 10 and 0.001, and 10 and 0.1, respectively, each with 40 equally spaced logarithmic steps. The mean NN $D_{SH}$ was calculated for each combination of the values. The results are shown in Figure \ref{fig:bandwidths_matrix_peri_q_nn}. The figure shows that varying the bandwidth of the perihelion distance does not influence the final statistics of the synthetic data significantly. In contrast,  the statistics are greatly influenced by the argument of perihelion bandwidth. At values as low as $h_q = 0.01$ the mean NN $D_{SH}$ distance is still above 0.07, provided that $h_\omega$ is above 1.0.

\begin{figure}
  \includegraphics[width=\linewidth]{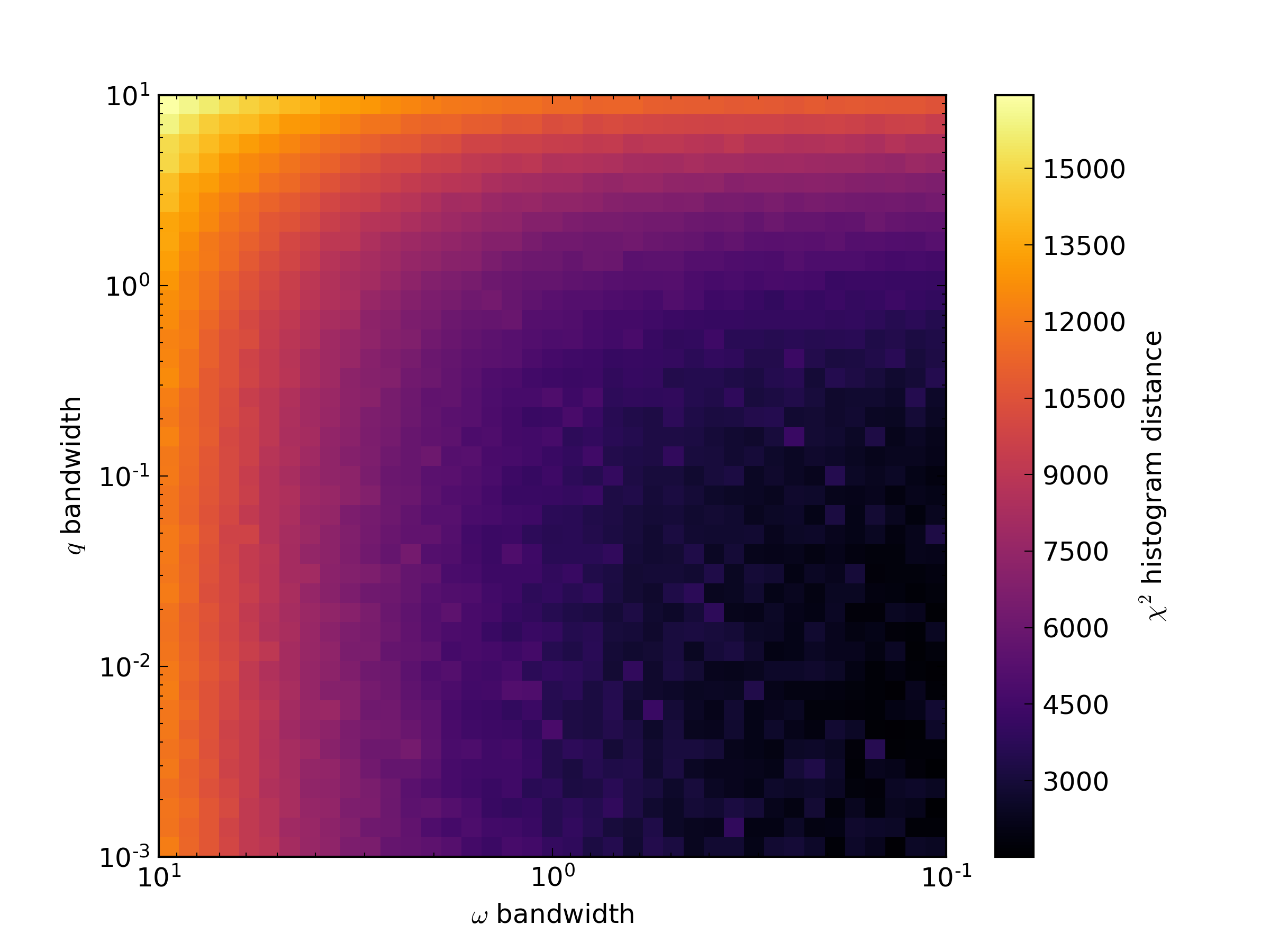}
  \caption{Influence of different values of $h_q$ and $h_\omega$ on the $\omega$ vs. $q$ $\chi^2$ histogram distance between the synthetic model and the observed dataset.}
  \label{fig:bandwidths_matrix_peri_q_histogram_diffs}
\end{figure}

A sense of how these values influence the final $\omega$ vs. $q$ histogram can be found by examining the $\chi^2$ histogram distance. The results, shown in Figure \ref{fig:bandwidths_matrix_peri_q_histogram_diffs} reveal a direct correlation of both $h_q$ and $h_\omega$. An optimal choice of bandwidth is one which preserves the data statistics and minimizes the $\chi^2$ histogram distance - thus values of $h_q = 0.01$ and $h_\omega = 6$ were chosen.

Based on the foregoing analysis, the final bandwidth matrix used was:

\begin{eqnarray}
\vec{H} = \begin{bmatrix}
  0.01     &      0        &      0     &        0     & 0 \\
   0       &     10        &      0     &        0     & 0 \\
   0       &      0        &     10     &        0     & 0 \\
   0       &      0        &      0     &       10     & 0\\
   0       &      0        &      0     &        0     & 6
\end{bmatrix}
\end{eqnarray}

\begin{figure}
  \includegraphics[width=\linewidth]{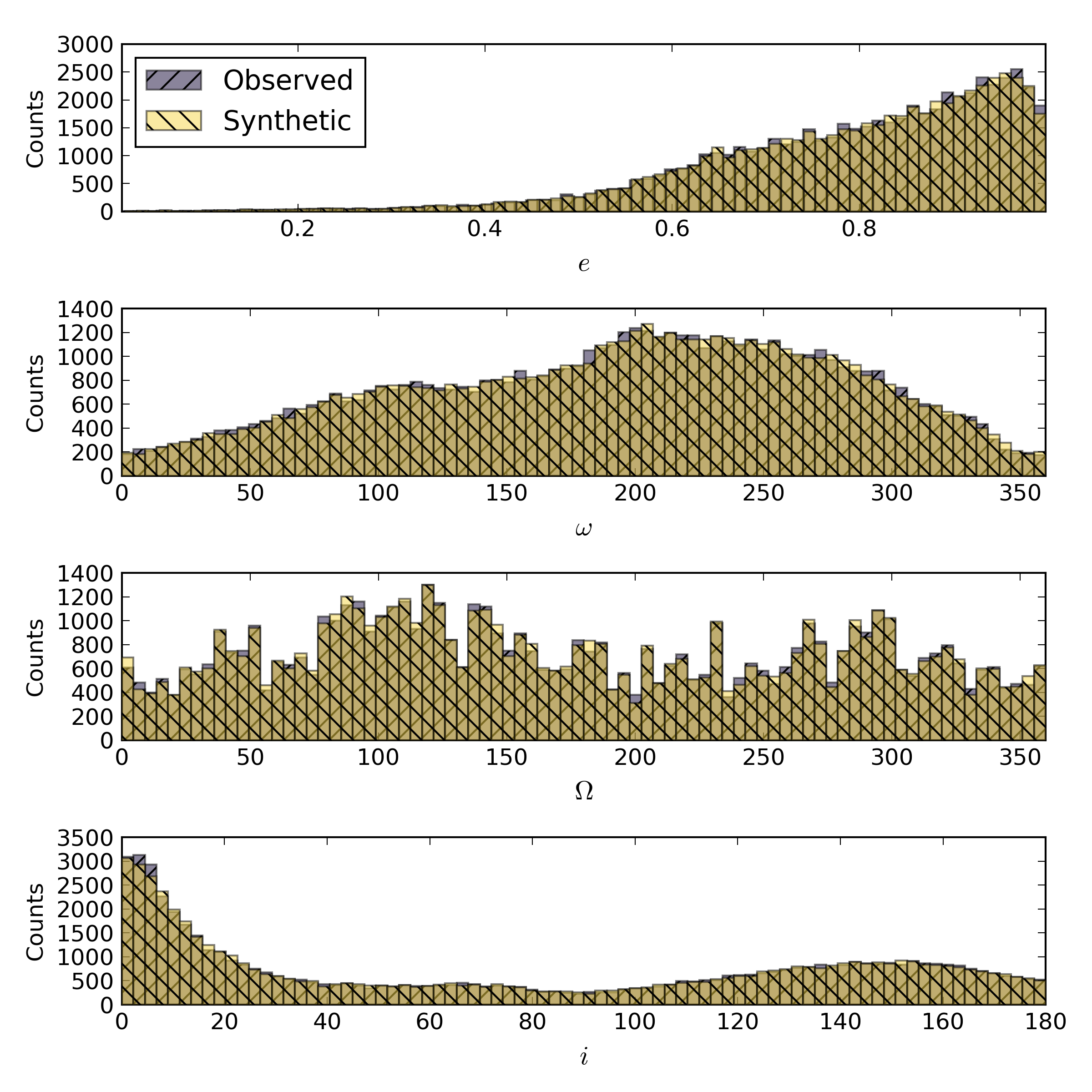}
  \caption{Comparison of histograms of all generated orbital parameters: $e$, $\omega$, $\Omega$ and $i$, using the non-scalar bandwidth matrix. Both the observed and synthetic orbits histograms are shown.}
  \label{fig:bandwidth_nonscalar_all}
\end{figure}

\begin{figure}
  \includegraphics[width=\linewidth]{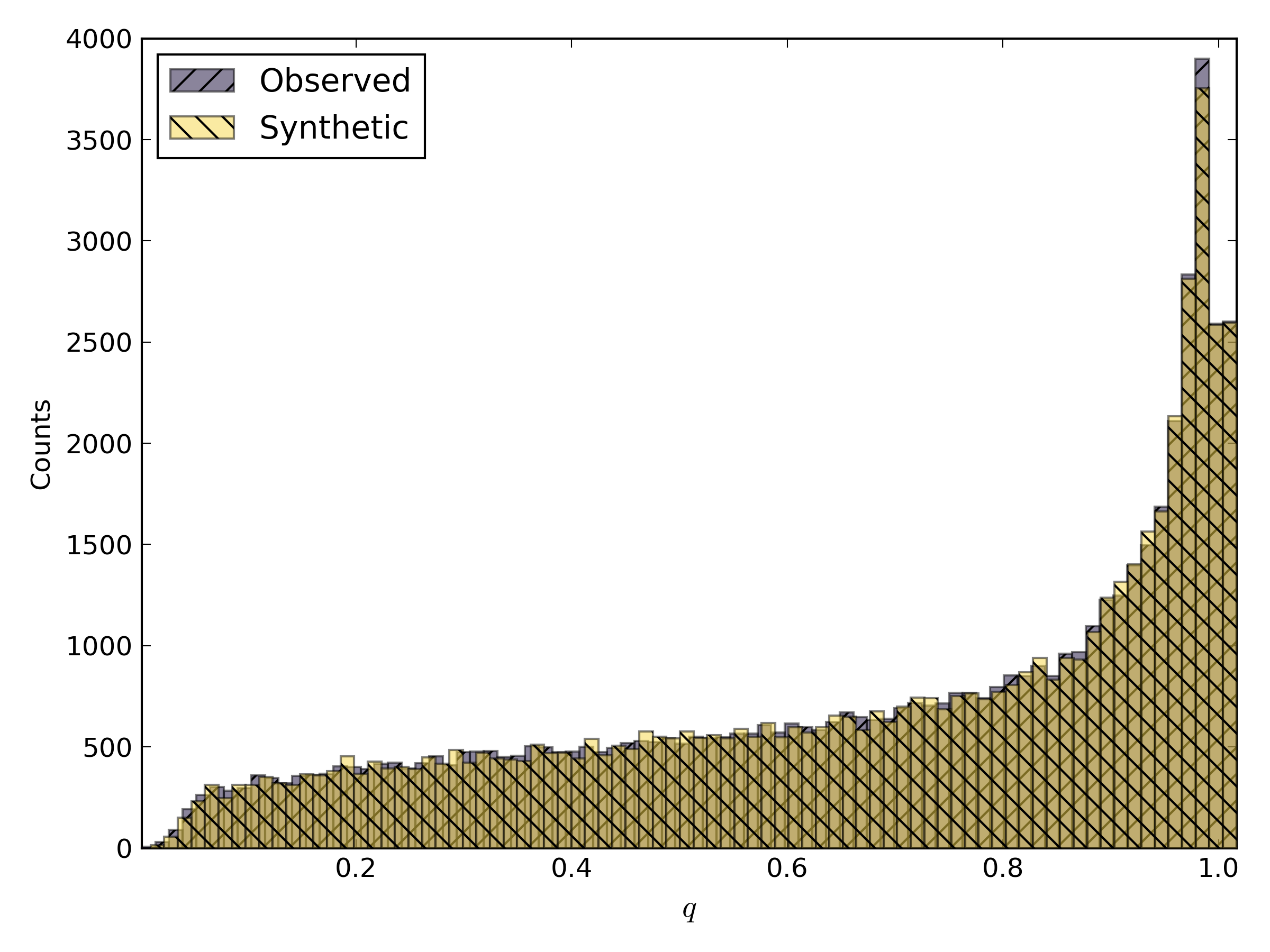}
  \caption{Perihelion distance histogram obtained by using the KDE with the non-scalar bandwidth matrix.}
  \label{fig:bandwidth_nonscalar_q}
\end{figure}

\begin{figure}
  \includegraphics[width=\linewidth]{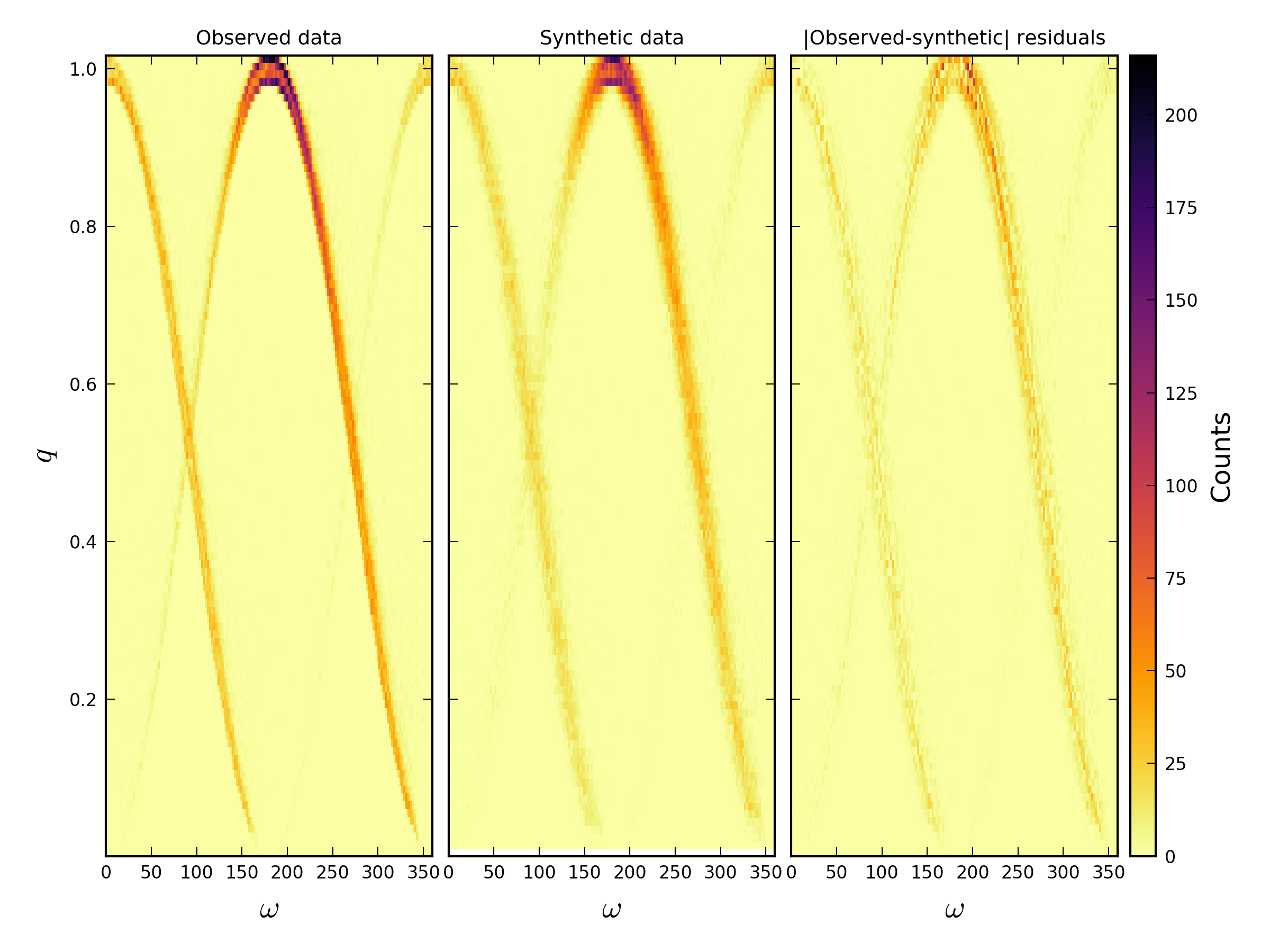}
  \caption{Argument of perihelion vs. perihelion distance for synthetic data obtained using the KDE with the non-scalar bandwidth matrix.}
  \label{fig:bandwidth_nonscalar_peri_vs_q}
\end{figure}

Selected results obtained with this bandwidth matrix are given in Figures \ref{fig:bandwidth_nonscalar_all} to \ref{fig:bandwidth_nonscalar_peri_vs_q}. The plots of other parameters show no visible difference from the scalar bandwidth matrix method.

Figure \ref{fig:bandwidth_nonscalar_all} shows single-parameter histograms of the synthetic data and compares them to the observed dataset. Compared to the $h=10$ synthetic data, shown in Figure \ref{fig:bandwidth_10_all}, the $\omega \approx 180\degree$ gap is no longer present, and all other data follow the original distribution.

The perihelion distance distribution, shown in Figure \ref{fig:bandwidth_nonscalar_q}, is now a very good reproduction of the observed data, comparable to the $h = 0.1$ results. Both the general trend in the histogram and the data range are consistent with the observed dataset.

Finally, Figure \ref{fig:bandwidth_nonscalar_peri_vs_q} shows the $\omega$ vs. $q$ density plots. Compared to the observed data, the synthetic plot is more dispersed, but still follows the observed data distribution. Both  main branches are now reproduced, and the central branch shows high and low $q$ branches. Furthermore, compared to the $h=10$ synthetic data (Figure \ref{fig:bandwidth_10_peri_vs_q}), the non-scalar bandwidth matrix data are less dispersed and show features matching the original dataset that are not present in the synthetic data using $h=10$.

Overall, we find that the non-scalar bandwidth matrix method has proven to be an optimal compromise between preserving the data structure and the underlying statistics. The mean NN $D_{SH}$ value for data generated with this method was 0.075, only $-6.8\%$ different from the value obtained for observed data.

We also manually explored the influence of changing the bandwidth of other parameters. In each case it was found that they produce worse fits in both the general statistics and the $\omega$ vs. $q$ histogram distance compared to the observed data. As independently changing individual bandwidth parameters produced poorer results in general, bandwidth values other than $h_q$ and $h_\omega$ were not changed.

\section{Discussion}
\subsection{Comparative analysis of results}
To quantify the diversity of among synthetically generated datasets, and as a check that the produced orbits were not too similar to their originals, each set of data and each synthetic orbit was paired with each observed orbit. The \cite{southworth1963statistics} D criterion was then calculated. The minimum value for each orbit was found and the values averaged. The results are presented in the first column of table \ref{tab:quantitative_analysis}. 

The orbits produced by the \cite{jopek2016probability} method E show the highest difference compared to observed orbits of all methods, at 0.0998. As this type of synthetic orbits has shown a large difference in orbital elements from observed orbits, we consider the difference of 0.0998 as an upper bound for an acceptable synthetic orbit dataset. This method also has the highest mean NN $D_{SH}$ value of 0.1014, almost 0.02 higher than that of the observed data. Furthermore, method E was found to have the largest difference in the $\omega$ vs. $q$ histogram from the observed data, compared to the methods tested here.

The kernel density estimation with the bandwidth of $h=10$ produced data that had orbits of equal statistical similarity as the observed data, the mean NN $D_{SH}$ being around 0.08. The drawback of this method was the high difference in the $\omega$ vs. $q$ histogram.

In contrast, data produced with $h=0.1$ produced histograms of high similarity to the observed data, but at the expense of a low mean NN $D_{SH}$ of 0.0425, half of the observed data value. This model also had the highest similarity to the observed orbits, of 0.0389 in $D_{SH}$; not a desirable feature. Thus, we consider the value of 0.0389 to be the lower bound for an acceptable synthetic orbit dataset.

Ultimately, to find a compromise between these competing factors, a non-scalar bandwidth matrix was used for the kernel density estimate. It produced synthetic data with an mean NN $D_{SH}$ of 0.075, only 0.0055 lower than the observed data, and a $\chi^2$ histogram distance, halfway between the $h=10$ and $h=0.1$ synthetic data. Moreover, the data had a mean difference from the observed data of 0.066 in $D_{SH}$, which is approximately half way between the set bounds for this value.

\begin{table*}[t] 
	\caption{Quantitative comparison of the observed and synthetically generated datasets.}
	\label{tab:quantitative_analysis} 
	\centering 
	\begin{tabular}{l c c c} 
	\hline\hline 
	Data & Mean minimum $D_{SH}$    & Mean nearest     & $\omega$ vs. $q$ histogram \\
	     & difference from observed & neighbour $D_{SH}$ & $\chi^2$ distance \\
	\hline 
	Observed & 0 & 0.0805 & 0\\
	Method E & 0.0998 & 0.1014 & 19712\\
	KDE, $h=10$ & 0.0730 & 0.0815 & 16456\\
	KDE, $h=0.1$ & 0.0389 & 0.0425 & 1954\\
	KDE, non-scalar& 0.0660 & 0.0750 & 8959 \\
	\hline 
	\end{tabular}
\end{table*}

\subsection{Method details}
 As the synthetic orbits were generated using a model which was built with a set of real meteor orbits, the results are highly dependant on the input data. The generated sporadic background will be only as good as the input data - if a meteor shower was not removed from the data, or it has a major observational bias, these will also be reflected in the final results. As a result, synthetic orbits generated with this method are only useful for comparison with real orbits obtained with the same system. Here our analysis used heliocentric orbital elements; nothing restricts the method to orbital elements; geocentric quantities could also be used as a distance metric (eg. \citet{valsecchi1999meteoroid}. One could also use the orbital state vector for meteoroids intersecting the Earth reducing the number of dimensions. One possible limitation of such an approach, however,  would be the problem of defining the covariance matrix of the state vector, though it might be possible to estimate the proper bandwidth values empirically.

A limitation of the approach presented in this work are the assumptions that all data  have the same uncertainty and bandwidth, and that the metrics between individual parameters are the same. While the former is somewhat alleviated by the fact that the orbits with high uncertainties were filtered out and they did not influence the estimation of the standard deviation, the latter presents a more complex issue as the orbital parameters are not strictly independent of each other, despite being scaled by their uncertainties.

Finally, although the analysis in this paper was limited to mimicking meteor data for generation of synthetic datasets, there are no obstacles to generalizing this to other Solar System objects. 

\section{Conclusion}
A new method for generating synthetic meteoroid orbits patterned after an existing dataset has been presented and tested. The method utilizes the Kernel Density Estimation (KDE) as a basis for building a data model from which synthetic orbits are drawn. The method produces estimates for all five Keplerian orbital parameters ($q$, $e$, $i$, $\Omega$, $\omega$) at once and preserves the underlying correlations between individual parameters inherent in the original observed data. A Gaussian kernel was used and its covariance matrix was assumed to be a diagonal matrix having elements equal to the squared mean standard deviation values of each observed orbital element. The CAMS orbit database was used as the seed database of orbits. During the investigation it was found that the results are heavily dependent on the choice of the KDE bandwidth matrix. 

First, a scalar bandwidth matrix was assumed and two scalar values of the bandwidth $h$ were explored: 10 and 0.1. The results were evaluated by comparing two values to ones obtained for the observed data: $\chi^2$ distance in the $\omega$ vs. $q$ density plots, and the mean nearest neighbour distance calculated using the \cite{southworth1963statistics} D criterion. With $h=10$ it was found that that the synthetic data reproduce the original value of the mean nearest neighbour distance of 0.08, but the difference in the density plot was high and the synthetic data did not reproduce the original structure. On the other hand, with $h=0.1$ if was determined that the $\omega$ vs. $q$ density plot is highly similar to the one obtained for the observed data, and the structure was preserved, but the mean nearest neighbour distance was half the observed data value, at 0.04. This is undesirable because it could lead to overestimating the number of orbits in dense areas of the parameter space, while underestimating the number of orbits in sparsely populated areas of the parameter space.

To find a compromise between these two evaluated parameters, a non-scalar bandwidth matrix was used. The bandwidths for $e$, $i$ and $\Omega$ parameters were $h_e = h_i = h_{\Omega} = 10$, the $q$ bandwidth was $h_q = 0.01$, and the $\omega$ bandwidth was $h_{\omega} = 6$. The results showed that the mean NN distance was close to the one obtained for the observed data, 0.075, while the $\omega$ vs. $q$ density plot preserved the data structure, although not with the same level of accuracy.

Finally, compared to the method E presented in \cite{jopek2016probability}, we found that the KDE with a non-scalar bandwidth matrix approach was the best of all methods of synthetic data evaluation attempted.

\subsection{Note on code availability and Python implementation of KDE approach} \label{implementation details}
A major goal in developing this technique is to allow reproducibility by others working on the specific problem of false-positive identification in meteoroid orbital datasets. As such, the method described above was implemented using the Python programming language, the data handled using the numpy library and the multivariate KDE implementation called from the scipy.stats library. The authors have created a Python module one can use to apply the method to any meteoroid orbit dataset. The module can be accessed on GitHub\footnote{Synthetic meteor orbit module on GitHub, \url{https://github.com/dvida/GenMeteorSporadics}}. The code is released under the MIT license. Readers are encouraged to contact the authors in the event they are not able to obtain the code on-line.

\section{Acknowledgements}
Funding for this work was provided through NASA co-operative agreement NNX15AC94A, the Natural Sciences and Engineering Research Council of Canada and the Canada Research Chairs Program. Furthermore, authors would like to thank Dr. Althea Moorhead for constructive comments, and Prof. Paul Wiegert for fruitful discussions and valuable suggestions. We also thank the anonymous reviewer and Dr. David Asher for excellent suggestions to this work.

\bibliography{bibliography}

\end{document}